\documentclass[aps,
		prd,
		reprint,
        superscriptaddress,
		  nofootinbib,
		floatfix,
		notitlepage, onecolumn
		]{revtex4-2}
\pdfoutput=1

\usepackage{amsmath}
\usepackage{mathtools}
\usepackage{amssymb}
\usepackage{slashed}
\usepackage{color}
\usepackage{xcolor}
\usepackage{hyperref}
\usepackage{braket}
\hypersetup{colorlinks=true,linkcolor=blue} 
\usepackage{comment}
\usepackage{scalerel}
\allowdisplaybreaks

\usepackage{feynmp-auto}
\usepackage{tikz}
\usetikzlibrary{decorations.pathmorphing,decorations.markings}
\tikzset{snake it/.style={decorate, decoration={snake,segment length=4}}}
\tikzset{smallsnake/.style={decorate, decoration={snake,segment length=2, amplitude=1}}}
\newcommand{\xnot}{x_{0}}

\newcommand{\be}{\begin{equation}}
\newcommand{\ee}{\end{equation}}
\newcommand{\bi}{\begin{enumerate}}
\newcommand{\ei}{\end{enumerate}}
\newcommand{\ud}{{\mathrm{d}}}
\hfuzz=\maxdimen \tolerance=10000
\vfuzz=\maxdimen \tolerance=10000
\newcommand{\LCm}{{\scriptscriptstyle -}}
\newcommand{\LCp}{{\scriptscriptstyle +}}

\newcommand{\LCperp}{{\scriptscriptstyle \perp}}
\renewcommand{\j}{\theta}
\newcommand{\nn}{\mathfrak{n}}

\newcommand{\tri}{ 
\mathchoice 
{ {\scalebox{0.6}{$\triangle$}} }
{ {\scalebox{0.575}{$\triangle$}} }
{ {\scalebox{0.55}{$\triangle$}} }
{ {\scalebox{0.4}{$\triangle$}} }
}

\newcommand{\llangle}{\langle\!\langle}
\newcommand{\rrangle}{\rangle\!\rangle}
\def\ddel{{}^\bullet\! \Delta}
\def\deld{\Delta^{\hskip -.5mm \bullet}}

\def\ddeld{{}^{\bullet}\! \Delta^{\hskip -.5mm \bullet}}

\def\Gdd{\ddot{G}}
\def\Gd{\dot{G}}

\newcommand{\rf}{\mathrm{f}}
\newcommand{\e}{\mathrm{e}}

\begin{document}

\title{Scattering in strong field QED in a non-null background}

\author{Patrick Copinger}
\email{copinger@hiroshima-u.ac.jp}
\affiliation{International Institute for Sustainability with Knotted Chiral Meta Matter (WPI-SKCM${}^2$), Hiroshima University, 1-3-1 Kagamiyama, Higashi-Hiroshima, Hiroshima 739-8531, Japan}
\affiliation{Centre for Mathematical Sciences, University of Plymouth, Plymouth, PL4 8AA, UK}
\author{James P. Edwards}
\email{james.p.edwards@plymouth.ac.uk}
\affiliation{Centre for Mathematical Sciences, University of Plymouth, Plymouth, PL4 8AA, UK}
\author{Karthik Rajeev}
\email{karthik.rajeev@ed.ac.uk}
\affiliation{Higgs Centre, School of Physics and Astronomy, University of Edinburgh, EH9 3FD, UK}

\begin{abstract}
\noindent We examine scattering amplitudes for an arbitrary number of photons in a class of non-null background electromagnetic fields, studying tree-level and one-loop amplitudes in scalar and spinor quantum-electrodynamics in backgrounds defined by a gauge field $A_\mu(\nn\cdot x)$ for $\nn^2\neq 0$. Motivated to account for more physically realistic laser-plasma dispersive properties, our approach overcomes prior work studying such amplitudes in a constant background field and relaxes the familiar null criterion assumed for plane waves. \\

\noindent Master Formulae for the $N$-photon amplitudes dressed by the non-null background are constructed using the first-quantised worldline formalism, which can systematically account for all orders in the non-null parameter, $\nn^2$, treated here as an expansion parameter. These are derived from worldline representations of the coordinate and momentum space propagators (and their LSZ-truncated amplitudes) and the effective action, each incorporating the non-null background non-perturbatively. We then outline a partial resummation of their expansions in $\nn^{2}$. A special exactly solvable case of non-null constant crossed fields without photon insertion in the effective action is explored to test the Master Formulae that result. The validity of the presented master formulae is further checked against known expressions for the wavefunction and non-linear Compton scattering in a non-null background to lowest order in the non-null parameter.
\end{abstract}

\maketitle

\onecolumngrid

\section{Introduction}

Theoretical studies of quantum field theory in electromagnetic backgrounds (i.e., strong-field quantum electrodynamics -- SFQED) have long relied on approximating the electromagnetic backgrounds as being some combination of constant and/or plane wave fields \cite{Fedotov:2022ely}, which have become central analytics tools. In such fields, the Dirac and Klein–Gordon equations admit closed-form (Volkov) solutions~\cite{Wolkow:1935zz, PhysRev.151.1058,osti_4032026,osti_4046383}, from which the field-dressed matter wavefunctions and propagator can be derived exactly. This allows standard S-matrix techniques to be applied to determine scattering amplitudes analytically (for arbitrary background profiles, these are integral representations) and non-perturbatively in the background coupling strength~\cite{goldman1964intensity,PhysRev.138.B740}. The earliest studies go back almost to the birth of quantum field theory itself, considering the quantum effective action \cite{Heisenberg:1936nmg, Weisskopf:1936hya}, light-by-light scattering \cite{Euler:1935zz, Akhieser, Karplus:1950zza, Karplus:1950zz, DeTollis:1964una, DeTollis:1965vna} and the Sauter-Schwinger pair creation effect 
\cite{Sauter, Schwinger} -- for a contemporary review of these developments see \cite{Scharnhorst:2017wzh, Fradkin:1991zq, Dittrich:2000zu, Ahmadiniaz:2024xob}. 

In particular, amplitudes in a plane-wave background, characterised by a gauge field of the form $A_{\mu}(n\cdot x)$ with $n^{2} = 0$ null, play a role in strong-field QED analogous to the hydrogen atom in quantum mechanics: they provide a rare, highly symmetric setting in which nonlinear interactions can be treated exactly, and they supply benchmark results against which approximations and numerical methods can be tested. Much of our present understanding of strong-field processes—nonlinear Compton scattering~\cite{Podszus:2021lms,Dinu:2013hsd}, nonlinear Breit–Wheeler pair creation~\cite{Meuren:2015mra}, and the nonlinear trident process~\cite{Tang:2022ixj} derives from this exceptional solubility.

More recently, the so-called worldline formalism of quantum field theory \cite{UsRep, ChrisRev} has been extended to incorporate constant fields \cite{Reuter:1996zm, Schmidt:1993rk, Shaisultanov:1995tm, Dunne:2002qf, Dunne:2002qg, Alexandrou:1998ia, Ahmad_2017, fppaper3} and plane wave backgrounds \cite{Edwards:2021vhg,Edwards:2021uif,Copinger:2023ctz, Schubert:2023gsl}. This first quantised approach, initially proposed in ~\cite{Feyn1,Feyn2,Strass1}, is known for its calculational efficiency, allowing derivation of ``Master Formulae'' that simultaneously represent amplitudes for an arbitrary number of scattering photons. It is especially well-adapted to studying higher-multiplicity scattering processes, because the Master Formulae turn out to combine multiple Feynman diagrams related by permutation of external legs \cite{Guzman:2026jfr, Schubert:2023bed, Edwards:2021uif}, thereby better representing the physical characteristics and gauge structure of the amplitudes. Its success is well-illustrated by the recent discovery of previously overlooked 1PR contributions to scattering amplitudes in homogeneous backgrounds \cite{Edwards:2017bte, Ahmadiniaz:2017rrk, Ahmadiniaz:2019nhk}, which it corroborated and generalised to tree-level processes. Much less has been done beyond this constant- or plane-wave field limit (but see \cite{Tarasov:2019rfp, Copinger:2024twl}).

In parallel to theoretical developments, modern high-intense laser facilities \cite{LUXE:2023crk, Clarke:2022rbd, PhysRevAccelBeams.22.101301,  Sacla, Corels, SEL} are pushing into regimes where the idealisation of the background field as a pure plane-wave vacuum-solution to the Maxwell equations may no longer suffice~\cite{PhysRevE.83.026406,Raicher:2013cja,DiPiazza:2013vra,Oertel:2015yma}. In realistic ultra-intensity laser experiments, the laser often interacts with residual gas, ionised matter, or even intentionally injected plasmas — leading to non-trivial medium responses such as dispersion, refractive index shifts~\cite{Cronstrom}, and collective effects~\cite{Mackenroth:2018rtp,Mackenroth:2020pct}. 
In particular, when propagating through a plasma, the phase-four vector describing the field wavefronts ceases to be strictly null, and the field no longer behaves as a classical plane wave in vacuum. As a result, the background vector potential can often be better modeled by a form akin to $eA_{\mu}(x) = \delta_{\mu}^{\perp}a_{\perp}(\nn \cdot x)$, with $\nn^2\neq 0$ even though the wavefronts remain planar~\cite{Cronstrom,Becker-non-null,Heinzl:2016kzb} and perpendicular to the light-front directions (see below); such gauge fields we refer to as \textit{non-null} fields. This departure from \textit{nullness}, in turn, encodes the medium’s dispersive properties (effective photon ``mass'', refractive index, group velocity effects, etc.), thereby capturing some of the realistic physics of laser–plasma interactions in strong-field regimes~\cite{King:2016xav,Dmitrieva:2025ohn}.

\begin{figure}
    \centering
    \begin{align*}
    \begin{tikzpicture}
   \begin{scope}[decoration={markings,mark=at    position 0.5 with {\arrow{>}}}] 
    \draw[postaction={decorate}, double] (-1,0)--(1,0);
   \end{scope}
    \end{tikzpicture}=
   \underbrace{\left(\sum_{N} \begin{tikzpicture}[baseline]
    \begin{scope}[decoration={markings,mark=at position 0.5 with {\arrow{>}}}] 
       \draw[postaction={decorate}] (-1,0)--(1,0);
       \draw[postaction={decorate},line width=0.5mm,red] (-1,.5)--(1,.5);
       \draw[draw=blue, smallsnake] (-.7,-0.9)--(-.7,0);
       \draw[draw=blue, smallsnake] (.7,-0.9)--(.7,0);     
       \node at (0,-0.9) {$...$};
       \node at (0,-.5) {$N$};
       \end{scope}
   \end{tikzpicture} \right)}_{\textrm{Plane wave resummation}}+ \underbrace{\left(\sum_{N} \begin{tikzpicture}[baseline]
    \begin{scope}[decoration={markings,mark=at position 0.5 with {\arrow{>}}}] 
    \draw[postaction={decorate}] (-1,0)--(1,0);
       \draw[postaction={decorate},line width=0.5mm,red] (-1,-0.9)--(-.7,-.7);
       \draw[line width=0.5mm,red] (-.7,-.7)--(-.5,-.7);
       \draw[postaction={decorate},line width=0.5mm,red] (-.5,-.7)--(0,-.5);
       \draw[blue, smallsnake] (-.7,-1)--(-.7,-.7);
       \draw[blue, smallsnake] (-.5,-.7)--(-.5,0);
       \draw[draw=blue,smallsnake] (.7,-0.9)--(.7,0); 
       \fill[gray] (-.6,-.7) circle (4.5pt);
       \node at (.25,-0.9) {$...$};
       \node at (.25,-.75) {$N$};
       \end{scope}
   \end{tikzpicture} +...\right)}_{\textrm{Dispersive corrections }\sim \mathcal{O}(\rho^2)}+...
    \end{align*}
   \caption{A schematic representation of the background propagator (black double-line) in a possible realisation of a non-null plane wave background, treated as plane waves in a dispersive medium. The thick red represents particles in the medium. The leading-order propagator, which ignores interactions between photons and medium particles, corresponds to the plane-wave propagator. Higher-order correlations, as shown here, result from the scattering of photons off the medium particles.}
   \label{fig:dispersion}
\end{figure}
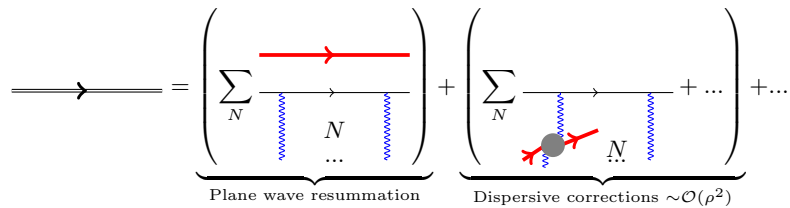

The exact analytic machinery of vacuum plane-wave QED, however, ceases to work in these non-null backgrounds. Our central idea is to exploit the fact that these backgrounds can still be viewed as controlled perturbations of a plane wave, with the non-nullness $\nn^2\equiv \rho^2$, serving as an expansion parameter. The plane wave-like part of the background can still be treated exactly (in this case, to all orders in $\rho^{2}$). An obvious advantage of this approach is that it preserves much of the analytic structure and clarity of the plane-wave case, while revealing how key physical quantities are modified once the null condition is relaxed. Physically, the approach can be thought of as retaining nonperturbative dressing of charged particles by the background through a Volkov-type exponentiation to ensure that part of the all-order interactions with the strong background is resummed~\cite{PhysRevE.83.026406,raicher2015novel,Raicher:2016bbx,Heinzl:2016kzb}, while incorporating effects such as medium-induced dispersion only at the perturbative level (see, Fig.~\ref{fig:dispersion}.)

From a methodological standpoint, our analysis is carried out using the worldline path integral formalism, which allows background-field effects to be incorporated in a transparent and often remarkably compact way~\cite{ChrisRev,UsRep,103}. In this approach, propagators, amplitudes, and other relevant objects are represented as path integrals over relativistic point particle trajectories. Over the past decade, the worldline formalism has seen significant development in the context of strong-field processes, including exact and near-exact treatments of plane-wave backgrounds~\cite{Schubert:2023gsl, Edwards:2021vhg,Edwards:2021uif,Copinger:2023ctz}, where it reproduces Volkov physics in a natural and economical manner. In particular, the integrability of the plane-wave background shows up in the worldline formalism as a hidden Gaussianity of the path integral, making it semi-classically exact (see \cite{Ilderton:2016qpj, Edwards:2021vhg}). These features make the worldline approach especially well suited for the present problem: it allows nonperturbative plane-wave effects to be treated exactly, while accommodating systematic perturbative insertions associated with deviations from null propagation.

More broadly, the framework we present here fits within a larger effort to extend strong-field QED to realistic electromagnetic backgrounds by exploiting the unique analytic control afforded by plane waves. In this perspective, the exact plane-wave amplitudes---already resumming the dominant nonperturbative dynamics --serve as an analytic core from which amplitudes in more general, non-plane-wave fields can be generated through systematic deformations. A recent example in this direction was taken in~\cite{Copinger:2024twl}, where amplitudes in pp-wave backgrounds were expressed as statistical averages over plane-wave ones. The present work represents another concrete step in this program -- systematically incorporating small deviations from the nullness of phase directions -- and motivates the long-term goal of extending the framework to fields with even more realistic features like transverse structure or finite spatial extent.

The organisation of this manuscript is as follows. In Section \ref{sect:non-null}, we introduce the class of non-null plane-wave backgrounds considered here, establish our conventions, and discuss their physical interpretation as controlled deformations of vacuum plane waves. Section \ref{sect:momentum-propagator} is devoted to scalar QED, where we develop the worldline representation of the $N$-photon–dressed propagator and effective action in non-null backgrounds, extending earlier plane-wave results to incorporate systematic corrections away from the null limit. In Section \ref{sect:spinor}, we conduct the corresponding analysis for spinor QED, highlighting both the similarities to and differences from the scalar case. We conclude in Section \ref{sect:conclusions} with a summary of our results and discuss possible extensions to more general backgrounds and phenomenological applications. Some lengthy calculations and comparison of our results with prior literature are included in Appendices.

\section{Non-Null Background Fields}
\label{sect:non-null}
Plane waves are solutions to Maxwell's equations characterised by wavefronts that are 3-dimensional planes. The normal to these planes is therefore $n\propto d\phi$, where $\phi\propto n\cdot x$ is the wavefront phase. The symmetry of this background suggests a convenient set of coordinates and a choice of gauge, whereby
the plane wave vector potential can be written as $eA_{\mu}(x) = a_{\mu}(n \cdot x) = \delta_{\mu}^{\perp} a_{\perp}(x^{+})$ with light-front direction $x^{+} = n \cdot x$ defined to align with the phase, the normal $n_{\mu}$ is null ($n^{2} = 0$), and the transversality to the direction of propagation of the plane wave is manifest (the ``$\perp$'' direction is orthogonal to $n$ so that $n \cdot a = 0$). Fields with such a gauge potential have a field strength tensor 
\begin{equation}
    \rf_{\mu\nu}({x^{+}}) = n_{\mu}a'_{\nu}(x^{+}) - n_{\nu}a'_{\mu}(x^{+})\,
\end{equation}
with field invariants
\begin{align}
    \mathcal{F}_{1} &= -\frac{1}{2}\rf_{\mu\nu}\rf^{\mu\nu} =  |\mathbf{E}|^{2} - |\mathbf{B}|^{2} = 0 \,, \\
     \mathcal{F}_{2} &= -\frac{1}{4}\rf_{\mu\nu}\widetilde{\rf}^{\mu\nu} = \mathbf{E} \cdot \mathbf{B} = 0\,,
\end{align}
where $\widetilde{\rf}^{\mu\nu} = \frac{1}{2}\epsilon^{\mu\nu\alpha\beta}\rf_{\alpha\beta}$ is the dual field strength tensor. So the electric and magnetic fields are orthogonal (``crossed'') and of equal magnitude (in natural units); with our convention, the fields point only in the transverse directions to the light-front coordinates. Such plane wave fields are relevant to SFQED because, as outlined in the introduction, the fields of high-intensity lasers are often approximately modeled as plane waves \cite{Wolkow:1935zz, Ritus:1985vta}; moreover, boosting an arbitrary electromagnetic field from the lab frame to the rest frame of an ultra-relativistic electron results in it seeing transverse fields significantly enhanced by the Lorentz factor, such that the electric and magnetic field are approximately perpendicular and of almost equal magnitude:
\begin{equation}
    \frac{\mathbf{E}' \cdot \mathbf{B}'}{|\mathbf{E}'||{\mathbf{B}'|}} \sim \frac{1}{2\gamma^{2}}  \frac{\mathbf{E} \cdot \mathbf{B}}{|\mathbf{E}||{\mathbf{B}|}}\sim 0 \,, \qquad \qquad \frac{|\mathbf{E}'|^{2} - |\mathbf{B}'|^{2}}{|\mathbf{E}'||{\mathbf{B}'|}} \sim \frac{1}{2\gamma^{2}}\frac{|\mathbf{E}|^{2}- |\mathbf{B}|^{2}}{|\mathbf{E}||{\mathbf{B}|}} \sim 0
\end{equation}
(where we are omitting dimensionless constants). One approximation regularly applied in QED calculations in very strong background fields is the ``Locally Constant Field Approximation'' (LCFA \cite{Baier:1998vh, osti_4032026, Ilderton:2018nws, PhysRevA.102.052805}) in which the (often exact) \textit{rates} for given processes in crossed constant fields (of equal magnitude) are integrated over the true, varying electromagnetic field profile. This models the process as taking place in a field that is instantaneously constant and generally works well when the variation of the fields is slow with respect to the Compton wavelength of particles participating in the process and its formation scale \cite{Blackburn:2018sfn, Aleksandrov:2019irn, PhysRevD.106.056013}. The LCFA is not, however, universally applicable and it is well known that it can miss important physical effects \cite{PhysRevA.91.013822, PhysRevA.98.012134, PhysRevLett.116.044801, PhysRevX.8.031004}.  

Although it is possible to add derivative corrections to the LCFA \cite{BAIER1989387, PhysRevA.101.042508, DiPiazza:2018bfu}, this does not always lead to a sensible expansion that reliably improves numerical results \cite{PhysRevA.99.042121}. Going beyond the plane-wave limit, here we instead examine electromagnetic waves whose wavefronts remain planar, but whose normal one-form $d\phi\propto\nn$ is no longer null ($\nn^2\neq0$). This means we can still choose to maintain $eA_{\mu}(x) = \delta_{\mu}^{\perp} a_{\perp}(x^{\tri})$ as transverse and depending on a single combination of coordinates, $x^{\tri} \equiv \nn \cdot x$, but relax the condition that this be a null direction. To describe this generalisation of a plane wave background field, it will be convenient to retain a reference null vector, $n \cdot a = 0$, defined such that in the limit $\nn^2\rightarrow 0$, we recover $\nn\rightarrow n$. This ensures that the background can be viewed as a well-defined deformation of a given plane wave. Our overall aim is then to extend the first-quantised worldline formulation of QFT to this class of backgrounds and to obtain Bern–Kosower–type master formulae for scattering amplitudes in these non-null background fields, building on existing developments for the plane-wave case.

To describe our deformation of the plane wave fields in a covariant way, it is useful to introduce a complementary null vector, $\bar{n}$ with $\bar{n}^{2} =0$, normalised such that $n\cdot \bar{n} = 2$. This allows us to define the corresponding light-cone coordinate $ x^{-}\equiv \bar{n}\cdot x $. For example, we may choose a coordinate system where $n_{\mu} = (1,0, 0, 1)$ and $\bar{n}_{\mu} = (1, 0, 0, -1)$. We can then express our non-null direction as $x^{\tri}= \nn \cdot x$, with
\begin{align}
    \nn_{\mu} &= n_{\mu} + \frac{\rho^{2}}{4}\bar{n}_{\mu}\,, \qquad \rho^2\equiv \nn^2\,,
\end{align}
which further leads to $\nn \cdot n = \rho^{2}/2$ and $\nn \cdot \bar{n} = 2$.
In the choice of coordinate system suggested above we then have $\nn_{\mu} = (1 + \rho^{2}/4, 0, 0, 1 - \rho^{2}/4)$. As such, $\rho^{2}$ (which can be positive or negative) serves as the deformation parameter away from a plane wave field, which satisfies $\lim_{\rho^2\rightarrow0}\nn_{\mu}(\rho^2)=n_{\mu}$. Note additionally that $\rho^2\rightarrow\infty$ also corresponds to a plane-wave limit, although the normal one-form for the wavefronts, in this limit, is proportional to $\bar{n}$ rather than $n$. Therefore, without loss of generality, we can restrict $-4\leq\rho^2\leq4$ and still cover the full space of deformations, as going outside of this range is equivalent to relabeling $n\longleftrightarrow\bar{n}$. Consequently, $|\rho^2/4|$ is bounded by unity, providing a well-defined deformation of the underlying plane wave. 

To illustrate the above characterisation of our deformed plane-wave backgrounds, consider, as an example, transverse plane waves in a homogeneous cold plasma. The Maxwell equation relevant to this case (see, e.g., \cite{chen2016introduction}) can be reduced to
\begin{align}
    \partial^2A_{\perp}-\Omega^2_{p}A_{\perp}=0\,, \quad n\cdot A=\bar{n}\cdot A=\partial^{\perp}A_{\perp}=0\,,
\end{align}
where $\Omega_{p}$ is the plasma frequency. The above equation can be solved exactly, for instance, by
\begin{align}\label{eq:plasma_nonnull}
    eA_{\perp}(x^{+},x^{-})=\frac{eE_0}{\Omega}\sin\left[\Omega\left(x^{+}+\frac{\Omega_{p}^2}{4\Omega^2}x^{-}\right)\right]\epsilon_{\perp}\,,
\end{align}
where $E_0$ is a constant and $\epsilon_{\perp}$ is a constant transverse polarisation vector. Comparing the above to the general form $eA_{\mu}(x) = \delta_{\mu}^{\perp} a_{\perp}(x^{\tri})$, we recognise that the field varies along a non-null direction of the same class, with the deformation parameter given explicitly by $\rho^2=\Omega^2_p/\Omega^2$. Then $\rho^2\ll1$ limit, in this context, translates to the case of the transverse wave's frequency $\Omega$, in the rest frame of the plasma, far exceeding the plasma-frequency $\Omega_p$. 

Denoting the phase-momentum of the background field as $\omega_{\mu} = \Omega\, \nn_{\mu}$, the parameter $\rho^2$ quantifies the \emph{smallness} of $\omega^2$. However, this notion of smallness is not unique, since the physical problem at hand may involve several other dimensionful parameters that can also set relevant scales. In the example above, denoting the four-velocity of plasma by $u^{\mu}$, the relevant smallness-condition may be stated covariantly as
\begin{align}\label{eq:small_rho_1}
    \frac{\omega^2}{(\omega\cdot u)^2}\sim\rho^2+\mathcal{O}(\rho^4)\ll 1\,.
\end{align}
However, as noted for example in \cite{PhysRevE.83.026406}, a careful analysis of the corrections to the Volkov wavefunction in such non-null fields leads to a less restrictive criterion \cite{Heinzl:2016kzb}, namely
\begin{align}\label{eq:small_rho_2}
    \frac{\omega^2}{\omega\cdot p}\ll 1\,,
\end{align}
where $p_{\mu}$ is the typical momentum of the charged probe particle, and in which case it is natural to choose $\frac{\omega^2}{\omega\cdot p}\sim \left(\frac{\Omega}{p^{\tri}}\right)\rho^2$ as the expansion parameter. Another possible small parameter emerges from an ultra-relativistic-type approximation in which
\begin{align}
  \frac{\omega^2 m^2}{(\omega\cdot p)^2}\ll 1\,,
\end{align}
corresponding to the expansion parameter $\frac{\omega^2 m^2}{(\omega\cdot p)^2}\sim \frac{m^2}{p^{+\,2}}\rho^2+\mathcal{O}(\rho^4)\,$.

In all of these cases, although the specific expansion parameters may vary, \(\rho^2\) conveniently serves as a bookkeeping parameter for the expansion. Therefore, to cover a wider range of physical situations within our class of non-null plane waves, we formally treat \(\rho^2\) as the expansion parameter; the correct physical scaling can always be reinstated at the end of the analysis through an appropriate rescaling of \(\rho^2\) (i.e., $\rho^2\rightarrow \left(\frac{\Omega}{p^{\tri}}\right)\rho^2$, $\rho^2\rightarrow \frac{m^2}{p^{+\,2}}\rho^2+\mathcal{O}(\rho^4)\,$, and so-on, as the case may be). This approach allows us, for example, to remain agnostic about the precise notion of smallness and instead focus on the strong-field QED consequences for scattering processes with arbitrary photon multiplicity in backgrounds of the form
\(
eA_{\mu}(x) = \delta_{\mu}^{\perp} a_{\perp}(x^{\tri})\,.
\)
We will also have the opportunity to provide a \textit{worldline-centric} perspective on what constitutes a small parameter and show how this relates to the physical, spacetime quantities discussed above.

\subsection{Non-null coordinates and fields}
\label{sec:coords}
Analogous to how light-front coordinates simplify calculations in a plane-wave background, aligning one coordinate with $x^{\tri}$ renders computations in the deformed plane-wave background considerably more tractable. Specifically, it is convenient to change the coordinates to
\begin{equation}
    (x^{+},x^{-},\boldsymbol{x}^{\perp})\rightarrow (x^{\tri}\,, x^{-}, \boldsymbol{x}^{\perp})\,,
\end{equation}
where the ``minus'' and transverse coordinates are carried over from the (limiting) plane-wave case (so $x^{-} = \bar{n}\cdot x$ and $\boldsymbol{x^{\perp}} = (x^{1}, x^{2})$). This coordinate choice offers a notable simplification in the worldline path-integral approach. As we present below, the kinematic term in the worldline action arises from the line-element, which in these coordinates is given by
\begin{equation}
    ds^{2} = dx^{\tri}dx^{-} - \frac{\rho^{2}}{4}(dx^{-})^{2} - dx^{\perp}\cdot dx^{\perp}\,,
    \label{eq:ds2}
\end{equation}
or equivalently, from the Minkowski metric written in this non-standard coordinate chart,
\begin{equation}
    \eta_{\mu\nu}=\frac{1}{2}(\nn_{\mu}\bar{n}_{\nu}+\nn_{\mu}\bar{n}_{\nu})-\frac{\rho^2}{4}\bar{n}_{\mu}\bar{n}_{\nu}-\sum_{i=1}^{2}(\epsilon^{\perp}_{i\mu}\epsilon^{\perp}_{i\nu})\,, 
\end{equation} 
where $\epsilon^{\perp}_{i}$ denotes the two orthonormal unit transverse vectors. As we will use these coordinates throughout, let us briefly examine how our characteristic vectors defining the electromagnetic field are represented in this coordinate system. The following components can be straightforwardly derived:
\begin{equation}
\begin{aligned}
    (\nn_{\tri},\nn_-,\nn_{\perp}) &=( 1\,, 0,0)\quad;\quad(\nn^{\tri},\nn^-,\nn^{\perp}) =( \rho^{2}\,, 2,0)\,,\\
    (\bar{n}_{\tri},\bar{n}_{-},\bar{n}_{\perp})&=(0,1,0)\quad;\quad(\bar{n}^{\tri},\bar{n}^{-},\bar{n}^{\perp})= (2, 0,0)\,,
\end{aligned}
\end{equation}
from which the non-zero components of the inverse-metric can be found to be $\eta^{\tri -}=\eta^{-\tri}=4$, $\eta^{\tri\tri}=\rho^2$ and $\eta^{ij}=-\delta^i_{j}$, where $i=1,2$ indicate the transverse coordinates. For an arbitrary four vector, $p$, with the contravariant and covariant components given respectively by $p^{\mu} = (p^{0}, p^{1}, p^{2}, p^{3})$ and $p_{\mu} = (p_{0}, p_{1}, p_{2}, p_{3})$ in the standard Minkowski coordinates, the corresponding components in the \emph{non-null} coordinate system are given in terms of the familiar components adapted to the null frame by
\begin{align}
    (p^{\tri},p^{-}) &= (\nn \cdot p,\bar{n}\cdot p) = \left(p^{+} + \frac{\rho^{2}}{4} p^{-},p^{0} - p^{3}\right)\,, \\
    (p_{\tri},p_{-}) &= \left(\frac{1}{2}p^-,\frac{1}{2}p^{\tri}-\frac{\rho^2}{4}p^-\right) = \left(p_{+},\frac{1}{2}p^{+} - \frac{\rho^{2}}{8}p^{-}\right)=\left(p_{+},\frac{1}{2}p^{+} - \frac{\rho^{2}}{4}p_{+}\right)\,,
\end{align}
where $p^{+} = n\cdot p$ and $p^{-} = \bar{n}\cdot p$ etc. The mass shell condition in the non-null coordinates then takes the form
\begin{equation}
    p^{\tri} =  \frac{\mathbf{p}^{2} + m^{2}}{p^{-}} + \frac{\rho^{2}}{4} p^{-}\,,
    \label{eq:MassShell}
\end{equation}
or indeed $p^{-} = \frac{2}{\rho^{2}} \Big(p^{\tri} - \sqrt{p^{\tri 2} - \rho^{2}(\mathbf{p}^{2} + m^{2})} \Big)$, being the solution with the correct limit as $\rho \to 0$.

With the coordinates and conventions established, we now return to a discussion of the background field. As mentioned above, the choice of dependence for the gauge potential ($eA_{\mu}(x) = \delta_{\mu}^{\perp} a_{\perp}(x^{\tri})$, absorbing the electromagnetic coupling to the background into $a_{\perp}$) leads to a one-parameter family of fields deforming the plane waves. The corresponding physical fields remain transverse, with the field strength tensor now taking the form
\begin{align}
    f_{\mu\nu}(x^{\tri}) &= \left(n_{\mu} + \frac{\rho^{2}}{4}\bar{n}_{\mu}\right)a'_{\nu}(x^{\tri}) -\left(n_{\nu} + \frac{\rho^{2}}{4}\bar{n}_{\nu}\right)a'_{\mu}(x^{\tri}) \\
    &=\nn_{\mu}a'_{\nu}(x^{\tri}) -\nn_{\nu}a'_{\mu}(x^{\tri})
\end{align}
As such the field invariants are now
\begin{align}\label{eq:F_1_def}
    \mathcal{F}_{1} &= -\rho^{2} a'^{2}\\
    \mathcal{F}_{2} &= 0\,,
\end{align}
so that the electric and magnetic fields remain orthogonal but are no longer of equal magnitude. In fact we can give closed formulae for arbitrary products of $f$ with itself:
\begin{align}
    \label{eq:feven}
    f^{2k}_{\mu\nu}  &=\big(\rho^{2} a_{\perp}'^{2}\big)^{k-1} f^{2}_{\mu\nu}\\
    f^{2k-1}_{\mu\nu} &=\big(\rho^{2}  a_{\perp}'^{2}\big)^{k-1}f_{\mu\nu}
    \label{eq:fodd}
\end{align}
where $f^{2}_{\mu\nu} = (n_{\mu} + \frac{\rho^{2}}{4}\bar{n}_{\mu})(n_{\nu} + \frac{\rho^{2}}{4}\bar{n}_{\nu})a_{\perp}'^{2} - \rho^{2}a'_{\mu}a'_{\nu}$.

At the level of the physical fields, in the transverse subspace, we have
\begin{align}
    \mathbf{E}_{\perp}(x^{\tri}) &= \big(1 + \frac{\rho^{2}}{4}\big)\big(a_{1}'(x^{\tri}), a_{2}'(x^{\tri})\big)\\
    \mathbf{B}_{\perp}(x^{\tri}) &= \big(1 -\frac{\rho^{2}}{4}\big)\big(a_{2}'(x^{\tri}), -a_{1}'(x^{\tri})\big)
\end{align}
so that 
\begin{equation}
   \frac{\mathbf{E} \cdot \mathbf{B}}{|\mathbf{E}||{\mathbf{B}|}} = 0\,, \qquad \qquad   \frac{|\mathbf{E}|^{2}-|\mathbf{B}|^{2}}{|\mathbf{E}||\mathbf{B}|} = \frac{\rho^{2}}{1 - \big(\frac{\rho^{2}}{4}\big)^{2}}\,.
\end{equation}
The second equality above establishes a field-centric interpretation of the deformation parameter $\rho^2$. Note that the expression on the right-hand side acts as a bijection in the range $-4\leq\rho^2\leq4$ which, as we already established, is enough to describe all possible deformations, modulo a coordinate change\footnote{We note in passing that the right-hand-side is invariant under $\rho^2\rightarrow-16\rho^{-2}$, under which $[-4,4]$ maps to $(-\infty,-4]\cup[4,\infty)$.}.
The fields are constant along lines of fixed $x^{\tri}$, pointing in the $(\bar{n}- \frac{\rho^{2}}{4} n)$-direction, and these can be spacelike ($\rho^{2} >0$) or timelike ($\rho^{2} < 0$). As deformations of the plane wave fields, we have
\begin{align}
    \begin{split}
        \mathbf{E}(x^{\tri}) &= \mathbf{E}_{0}(x^{+}) + \frac{\rho^{2}}{4}\big(\mathbf{E}_{0}(x^{+}) + x^{-} \mathbf{E}_{0}'(x^{+})\big) + \mathcal{O}(\rho^{4})\,\\
        \mathbf{B}(x^{\tri}) &= \mathbf{B}_{0}(x^{+}) - \frac{\rho^{2}}{4}\big( \mathbf{B}_{0}(x^{+}) - x^{-}\mathbf{B}_{0}'(x^{+}) \big)+ \mathcal{O}(\rho^{4})\,,
    \end{split}
    \label{eq:EBexp}
\end{align}
where the fields labeled with argument $x^{+}$ and subscript zero are exactly those that would appear in the plane wave case. Hence, we can also interpret our non-null parameter, $\rho^{2}$, as characterising the mixing of the two light-front directions in the electromagnetic fields and their difference in squared magnitude. In this sense, the chosen coordinate dependence of the field allows us to relax the strict vanishing of field invariants assumed by plane waves by allowing a non-zero difference in the fields' magnitudes and include all orders of derivative corrections relative to the constant field case (within the chosen family of fields). Let us also point out that the fields above establish that our deformation does not just include higher-derivative corrections but also a variation in the zeroth-order fields in order to satisfy the non-vanishing field invariant.

To illustrate this, we give the example, which we will return to again below, of constant crossed fields. For a plane wave such fields are described by $a_{\mu}(x^{+}) = a_{0} \delta_{\mu \perp} x^{+}$ so that with respect to a standard coordinate system $\mathbf{E} = a_{0}(1,1,0)$ and $\mathbf{B} = a_{0}(1, -1, 0)$ (constant, orthogonal fields of equal magnitude, which is the $\rho^{2} \to 0$ limit of \eqref{eq:EBexp}). The natural extension to the non-null case is for the gauge potential to be a linear function of $x^{\tri}$. Such a field arises, for instance, in the $\Omega\rightarrow 0$ limit (keeping $\rho^2=\Omega_p^2/\Omega^2$ fixed) of \eqref{eq:plasma_nonnull}. Alternatively, writing $a_{\mu}(x^{\tri}) = a_{0}\delta_{\mu\perp} x^{\tri}=a_{0}\delta_{\mu\perp} x^{+}+\rho^2/4 a_{0}\delta_{\mu\perp} x^{-}$, the same potential can also be thought of as describing the field of two counter propagating plane waves in the CCF limit. So with $a_{\mu}(x^{\tri}) = a_{0}\delta_{\mu\perp} x^{\tri}$ we have the (constant) fields
\begin{align}
\begin{split}
    \mathbf{E} &= a_{0}\big(1 + \frac{\rho^{2}}{4}\big)(1,1,0)\,\\
    \mathbf{B} &= a_{0}\big(1 - \frac{\rho^{2}}{4}\big)(1,-1,0)\,.
\end{split}
\label{eq:EBCrossed}
\end{align}
This deviation from the CCF has important physical consequences -- for instance, unlike the null plane wave, the non-null field can generate pairs via the Schwinger effect, which manifests, for example, in a nonzero imaginary part of the effective action (see below). 

Before proceeding, let us note that the above discussion presented the fields in a fixed but arbitrary reference frame. However, it is known  \cite{Heinzl:2016kzb, King:2016xav} that, depending on the sign of $\rho^{2}$, it may be convenient to Lorentz transform to a frame in which the field takes a specific form. Consider first the case that $\rho^{2} >0$: then there is a Lorentz boost along the $x^{3}$ direction (boost parameter $\gamma = \frac{1 + \frac{\rho^{2}}{4}}{\rho}$) such that the characteristic non-null direction is purely timelike, $\nn'_\mu = (\rho, 0, 0, 0)$. The electromagnetic field is then a spatially homogeneous but time-dependent electric field (``electric case''). If, on the other hand, $\rho^{2} < 0$ there is a Lorentz transformation (along $x^{3}$, $\gamma = \frac{1 - \frac{\rho^{2}}{4}}{|\rho|}$) such that $\nn'_\mu = (0, 0, 0, \rho)$ is purely spacelike and the electromagnetic field is static but spatially varying magnetic field (``magnetic case'')\footnote{The field for $\rho^{2} >0$ is in the same Lorentz class of fields as that of a plane wave propagating through a medium with refractive index $\textrm{n} < 1$; for $\rho^{2} < 0$ we are in the class in which belongs a plane wave traveling in a medium with refractive index $\textrm{n} > 1$ or that of a magnetic undulator/wiggler. For a more detailed discussion of charge dynamics in these fields, see \cite{Heinzl:2016kzb}.}. This classification will be useful when we discuss the Schwinger effect in our non-null fields below.

In the following section, we present a systematic framework for analysing such effects in our class of non-null deformed plane waves, by examining how the photon-dressed amplitudes are modified at both tree and loop level. We do this by forming the worldline representations of the matter propagator and the one-loop effective action dressed by both the non-null background field and external scattering photons.

\section{Scalar QED in Non-Null Backgrounds}\label{sect:momentum-propagator}
%

Having established the class of non-null plane-wave backgrounds and fixed our conventions, we now turn to the analysis of scattering processes in such fields. As outlined in the Introduction, our primary tool will be the worldline path-integral formalism, which provides a unified framework for treating both propagators and amplitudes in background-field QED, and treats scalar and spinor matter fields in a more similar way. In this section, we focus on scalar QED and we begin by reviewing the worldline approach to calculating the ``$N$-photon-dressed propagator'' and ``$N$-photon-dressed effective action'' in the presence of an additional semi-classical electromagnetic background defined by the vector potential $eA_{\mu}(x) = a_{\mu}(x)$. The photon-dressed propagator is the correlation function involving two scalar field insertions and $N$ (already LSZ-truncated) photon lines and upon LSZ reduction yields the tree-level $N+2$ particle scattering amplitude describing the interaction of $N$-photons with two external matter legs. The effective action corresponds to the one-loop $N$-particle scattering amplitude of the $N$-photons mediated by a virtual matter loop. In \cite{Copinger:2023ctz}, building on \cite{Ahmadiniaz:2021gsd, Ahmadiniaz:2020wlm, Ahmad:2016vvw} (respectively \cite{Schubert:2023gsl, Edwards:2021uif, Edwards:2021vhg}, building on \cite{Dittrich:2000wz, Schmidt:1993rk}), worldline representations for these objects in an exactly plane wave background were derived. In particular, the condition $n^{2} = 0$ implied particular simplifications that allowed for the analytic calculation of the worldline path integral. Here we return to the derivation of this result and show how the method can be systematically adapted to go beyond the null-field limit.

\subsection{N-photon dressed propagator and effective action}

The worldline representation of the matter propagator in scalar QED, as presented in \cite{Copinger:2023ctz}, is the following path integral over open worldlines propagating between fixed endpoints in arbitrary proper time:
\be\label{eq:propagator_pos_pi}
    \mathcal{D}^{x'x}[A] = 
    \int_{0}^{\infty}\ud T \,\e^{-im^{2} T}\,
\int^{x(T)=x'}_{x(0)=x}\hspace{-1.5em}\mathcal{D}x(\tau)\, \e^{iS[x(\tau),A]}\;,
\ee
where $S[x(\tau),A]$ denotes the action for a particle worldline, $x^\mu(\tau)$, propagating in propertime $T$, minimally coupled to the gauge field $A_{\mu}$, and is given by
\be
    S[x(\tau),A]= -\int_{0}^{T}\!\ud\tau\,\Big[\frac{\dot{x}^{2}}{4}+e{A(x(\tau))\cdot \dot{x}(\tau)} \Big]\;,
\ee
with dots denoting propertime differentiation, i.e. $\dot{x}(\tau)\coloneqq \tfrac{d}{d\tau} x(\tau)$. Similarly, as in \cite{Edwards:2021vhg}, a worldline representation of the one-loop effective action (here adapted to Minkowski signature) involves a path integral over closed worldlines 
\be\label{eq:gamma}
    \Gamma[A] = 
    -i\int_{0}^{\infty}\frac{\ud T}{T} \,\e^{-im^{2} T}\,
    \int_{x(0) = x(T)} \hspace{-1.5em}\mathcal{D}x(\tau)\, 
    \e^{iS[x(\tau),A]}\,.
\ee
Equations \eqref{eq:propagator_pos_pi}–\eqref{eq:gamma} make clear that the worldline formalism naturally accommodates both open and closed trajectories, depending on whether one is considering tree-or loop-level objects (and the formalism extends to higher loop order \cite{UsRep, ChrisRev, Fliegner:1997ra, Dunne:2002qf}). Moreover, this structure allows us to systematically incorporate background fields and scattering photons by splitting the gauge potential and treating interactions in a controlled expansion, as follows.

In our cases of interest, the gauge potential, $A_{\mu}(x)$ is decomposed as $eA_{\mu}(x) = a_{\mu}(x^{\tri})+ eA_{\mu}^{\gamma}(x)$, in which $a_{\mu}$ is the non-null plane wave background field as defined in Section \ref{sect:non-null}, for which the coordinate dependence enters solely through $\nn \cdot x=x^{\tri}$ and $\nn \cdot a = 0$,  and $A_{\mu}^{\gamma}(x) = \sum_{i= 1}^{N} \varepsilon_{i \mu} \e^{i k_{i} \cdot x}$ represents the external (scattering) photons. The interaction with the scattering photons is then treated by expanding to multi-linear order in the photon polarisation vectors, leading to the $N$-photon dressed propagator and effective action in the background:
\begin{align}
     \label{eq:propagator_posN}
    \mathcal{D}^{x'x}_N[a] &= 
    (-ie)^{N}
    \int_{0}^{\infty}\ud T \,\e^{-im^{2} T}\,
    \int^{x(T)=x'}_{x(0)=x}\hspace{-1.5em}\mathcal{D}x(\tau)\, 
    \e^{iS[x(\tau),A]}
    \prod_{i=1}^{N} V^{x'x}[\varepsilon_{i}, k_i]\;,\\
    \label{eq:gammaN}\Gamma_{N}[a] &= -i
    (-ie)^{N}
    \int_{0}^{\infty}\frac{\ud T}{T} \,\e^{-im^{2} T}\,
    \int_{x(0) = x(T)}\hspace{-1.5em}\mathcal{D}x(\tau)\,
    \e^{iS[x(\tau),A]}
    \prod_{i=1}^{N} V^{x'x}[\varepsilon_{i}, k_i]\;,
\end{align}
where the scattering photons are represented by the vertex operator (as in string theory)
\begin{equation}
    V^{x'x}[\varepsilon,k]\coloneqq \int_{0}^{T}d\tau\,\varepsilon\cdot\dot{x}(\tau)\, \e^{ik\cdot x(\tau)}\,.
\end{equation}
The presence of the vertex operators seems to spoil the otherwise standard structure of the path integral. To return to a more amenable path integral, we first exponentiate the prefactors in the vertex operators as 
\begin{equation}
    V^{x'x}[\varepsilon, k] = \int_{0}^{T} d\tau \, \e^{i k \cdot x(\tau) + \varepsilon \cdot \dot{x}(\tau)} \Big|_{\textrm{lin}\,N}\,\,,
\end{equation} 
where the instruction $\textrm{lin}\,N$ simply means that the result should be projected onto the contribution linear in $\varepsilon$. From a path integral viewpoint, this exponentiated representation incorporates the interaction with the $N$-photons simply by adding a linear source term $-\int_{0}^{T}\mathcal{J}_{\rm (0)}(\tau)\cdot x(\tau)d\tau$, directly into the action, where
\begin{align}
    \mathcal{J}^{\mu}_{\rm(0)}(\tau)\equiv i\sum_{i=1}^{N}\big(ik_{i}^\mu - \varepsilon_{i}^\mu\frac{\ud}{\ud\tau}\big)\delta(\tau-\tau_{i})
\end{align}
is supported at the insertion points of the photons along the worldline. Next, to facilitate the integral over $x(\tau)$, it is convenient to expand the worldline trajectories as $x(\tau)=x_{\rm cl.}(\tau)+q(\tau)$, where $x_{\rm cl.}(\tau)$ is the classical solution to the worldline equations of motion and boundary conditions in \textit{vacuum}, specifically
\begin{align}
    \frac{1}{2}\ddot{x}_{\rm cl.}^{\mu}(\tau)=\mathcal{J}^{\mu}_{\rm (0)}(\tau)\quad\&\quad \begin{cases}
        x_{\rm cl.}(0)=x,x(T)=x'\quad&;\quad \textrm{(Line)}\,,\\
        x_{\rm cl.}(0)=x(T)=\xnot\quad&;\quad \textrm{(Loop)}\,.
    \end{cases}
\end{align}
For the open line, this classical solution is a straight line between the endpoints, and for the loop, we choose the solution corresponding to expansion about the loop centre of mass.

Finally, we note that for arbitrary gauge potential, $a_{\mu}(x^{\tri})$, the path integral is not, apparently, Gaussian in $q^{\tri}$. In order to deal with the functional integral over $q^{\tri}$, we follow \cite{Schubert:2023gsl} and introduce two auxiliary worldline fields $\xi(\tau)$ and $\chi(\tau)$ such that the non-trivial interaction term can be rewritten as
\begin{equation}
 \e^{-i\int \ud\tau\, a(x_{\rm cl.}^{\tri}(\tau)+q^{\tri}(\tau)) \cdot \dot{q}(\tau)}  = \int \mathcal{D}\xi(\tau)\mathcal{D}\chi(\tau)\,
\e^{i\int \ud\tau\, \big[\chi(\tau)(\xi(\tau)-q^{\tri}(\tau))-a(x_{\rm cl.}^{\tri}(\tau)+\xi(\tau))\cdot\dot{q}(\tau)\big]}\,.
\end{equation}
The point is that the path integral over $q(\tau)$ is then Gaussian, so can be computed analytically. We now turn to the details of this calculation, beginning with the space-time representation of the $N$-photon dressed propagator.

\subsubsection{$N$-photon dressed propagator: space-time representation}
For the open line, we have Dirichlet boundary conditions (DBC) on $q(\tau)$, and we consider the following part of the path integral in $q(\tau)$:
\begin{equation}\label{eq:DBC_q_int}
    \int_{DBC} \hspace{-1em} \mathcal{D}q(\tau)\,
 \e^{i\int_{0}^{T}\ud\tau\big[-\frac{\dot{q}^{2}}{4}-\delta\mathcal{J}\cdot q\big]} = {-i(4\pi T)^{-2}}\,
 \exp\Big[
 -i\int_{0}^{T}\ud \tau \ud \tau'\,  \delta\mathcal{J}_\mu(\tau)\Delta(\tau, \tau')\delta\mathcal{J}^\mu(\tau')
 \Big] \;,
\end{equation}
with current $ \delta\mathcal{J}^\mu (\tau) := a^\mu(x_{\rm cl.}^{\tri}(\tau) + \xi(\tau))\frac{\ud}{\ud\tau}+\chi(\tau)\nn^\mu$ and worldline Green function 
\begin{equation}\label{eq:Delta}
    \Delta_{ij} \coloneqq \Delta(\tau_{i},\tau_{j}) = \frac{1}{2}|\tau_{i}-\tau_{j}|-\frac{1}{2}(\tau_{i}+\tau_{j})+\frac{\tau_{i}\tau_{j}}{T}\,.
\end{equation}
We have singled out the term in \eqref{eq:DBC_q_int}, because -- apart from the factor $\exp(i\int_{0}^{T} \xi(\tau) \chi(\tau) d\tau)$ -- it is the only other contribution that carries $\chi$ dependence and is therefore relevant for the subsequent integration over $\chi$. However, additional terms arise in the evaluation of \eqref{eq:propagator_posN}. First, because the path integral is expanded about the classical trajectory $x_{\rm cl.}(\tau)$, we acquire an analogous contribution to \eqref{eq:DBC_q_int} with $\delta\mathcal{J}$ replaced by $\mathcal{J}_{\rm (0)}$ (as discussed above this encodes the interaction with the scattering photons) which originates from evaluating the vacuum limit of the worldline action on the classical trajectory $x_{\rm cl.}(\tau)$. In this light, $\delta\mathcal{J}^\mu (\tau)$ plays an analogous role, acting as an effective current that captures the interaction with the non-null background field. Second, there's also a term in the exponential that is proportional to $\int_{0}^{T} d\tau\, a(x_{\rm cl.}^{\tri}(\tau) + \xi(\tau))\cdot \dot{x}_{\rm cl.}(\tau)$, which similarly arises from expanding the action about the classical path. These are the usual contributions one would obtain from the worldline path integral, well-treated in previous literature in a plane wave background \cite{Edwards:2021vhg,Edwards:2021uif,Copinger:2023ctz}, which are simply appended to \eqref{eq:DBC_q_int}.

It remains to integrate over $\chi(\tau)$ and $\xi(\tau)$, as in \cite{Schubert:2023gsl, Copinger:2023ctz}, though, here we relax the condition that $\rho^{2} = 0$. To this end, we return to the argument of the exponential in \eqref{eq:DBC_q_int}, which expands to
\begin{align}
\int_{0}^{T} d\tau d\tau'\, \delta\mathcal{J}(\tau)_{\mu} \Delta(\tau, \tau') \delta\mathcal{J}^{\mu} & =
\int_{0}^{T}\!\ud \tau \ud\tau' \, a(x_{\rm cl.}^{\tri}(\tau) + \xi(\tau))\cdot a(x_{\rm cl.}^{\tri}(\tau') + \xi(\tau'))\ddeld(\tau, \tau')+\rho^2 \int_{0}^{T}\!\ud \tau \ud\tau' \, \chi(\tau)\chi(\tau') \Delta(\tau, \tau')
 \label{eq:JDelJ}
\end{align}
in which $\ddel_{ij} := \frac{\partial}{\partial \tau_{i}} \Delta_{ij}$, $\deld_{ij} := \frac{\partial}{\partial \tau_{j}} \Delta_{ij}$ etc -- notation defined in \cite{103}. Note that the first term in the above equation combines with the previously mentioned $\chi$-independent contributions to reproduce essentially the plane-wave result, but with the replacement $a(x^{+}_{\rm cl.}(\tau)) \rightarrow a(x^{\tri}_{\rm cl.}(\tau)+\xi(\tau))$. This replacement occurs repeatedly in what follows, making it central to the structure of the deformation from nullness, and thus it is convenient to introduce a compact notation. Given a functional $\mathcal{F}[a]=\mathcal{F}[a(x_{\rm cl.}^{+}(\tau))]$, we denote 
\begin{align}\label{eq:def_replacement}
    \mathcal{F}[a(\bullet+\xi)]\equiv \mathcal{F}[a(x_{\rm cl.}^{\tri}(\tau)+\xi(\tau))]\,.
\end{align}
By contrast, the second term in \eqref{eq:JDelJ} constitutes a genuinely non-trivial correction due to the deviation from nullness, and, as we shall see shortly, the above replacement of the argument to the gauge field will play a key role in handling these corrections.

One may directly evaluate the path integral\footnote{For a different, more direct evaluation of this path integral, see Appendix~\ref{app:chi_integral}.} over $\chi(\tau)$; however to facilitate computing the integral over $\xi(\tau)$ (note that this variable enters the path integral inside the argument of an arbitrary gauge field) it is convenient to rewrite the path integral over $\chi(\tau)$
as (integrals in the exponent run from $\tau, \tau' = 0$ to $\tau, \tau' = T$)
\begin{equation}\label{eq:int_chi}
    \int \mathcal{D} \chi(\tau) \, \e^{-i \rho^{2}\int\!\ud \tau \ud\tau' \, \chi(\tau)\chi(\tau') \Delta(\tau, \tau') + i \int \ud\tau \chi(\tau) \xi(\tau)}=
    \e^{i \rho^{2}\int \ud \tau' \ud \tau'' \Delta(\tau', \tau'') \frac{\delta}{\delta \xi(\tau')} \frac{\delta}{\delta \xi(\tau'')}}\delta[\mathcal \xi(\tau)]  \,.
\end{equation}
Then, we may integrate by parts under the path integral to shift the functional derivatives onto the remaining terms in the integrand. This frees up the functional $\delta$-function, and it is easy to see that integrating over $\xi(\tau)$ then effects the subsequent replacement 
\begin{equation}\label{implicit_a_line}
    a_{\mu}\big(x^{\tri}_{\rm cl.}(\tau) + \xi(\tau)\big) \longrightarrow a_{\mu}(\tau) \coloneqq a_{\mu}\big(x^{\tri}_{\rm cl.}(\tau)\big) \,,
\end{equation}
where
\begin{align}
    x^{\mu}_{\rm cl.}(\tau)=x^{\mu} + z^{\mu}\frac{\tau}{T} + 2i\sum_{j=1}^{N} [\deld(\tau, \tau_{j}) \varepsilon_{j}^{\mu} + i\Delta(\tau, \tau_{j})k_{j}^{\mu}]\,,
\end{align}
and $z^{\mu}=x'^{\mu}-x^{\mu}$. The path integration has been formally completed, resulting in
\begin{align}
    & \mathcal{D}_{N}^{x'x}[a] = i(-ie)^{N}\int_{0}^{\infty} dT \, (4\pi iT)^{-2}\int_{0}^{T}\prod_{i=1}^{N}d\tau_i\,  \e^{i \rho^{2}\int \ud \tau' \ud \tau'' \Delta(\tau', \tau'') \frac{\delta}{\delta \xi(\tau')} \frac{\delta}{\delta \xi(\tau'')}}\e^{i\mathcal{S}^{\rm tr.}_{\rm pw.}[a(\bullet+\xi)]} \Big|_{\mathrm{lin}\,N}^{\xi(\tau) \rightarrow 0}\,,
\label{Dxx_full}
\end{align}
where $a(\bullet+\xi)$ indicates the replacement rule introduced in \eqref{eq:def_replacement}, and after straightforward evaluation, the on-shell action $\mathcal{S}^{\rm tr.}_{\rm pw.}[a]$ is, in fact, the corresponding exponential for the (null) plane wave background at tree level, given explicitly by

\pagebreak
\begin{align}
    i\mathcal{S}^{\rm tr.}_{\rm pw.}[a]= &-iM^{2}(a)T-i \frac{z^{2}}{4T} - i z \cdot \llangle a \rrangle + i\sum_{j=1}^{N} \left[ k_{j} \cdot (x + z \frac{\tau_{j}}{T}) - i \varepsilon_{j} \cdot \frac{z}{T} \right]\\\nonumber
    &+ 2\sum_{j=1}^{N} \big[ \big(\llangle a \rrangle - a(\tau_{j})\big)\cdot \varepsilon_{j} - i I(\tau_{j})\cdot k_{j} \big]
-i\sum_{i,j=1}^{N}[\Delta_{ij}k_{i}\cdot k_{j} - 2i\ddel_{ij}\varepsilon_{i}\cdot k_{j} - \ddeld_{ij} \varepsilon_{i}\cdot\varepsilon_{j}]\,,
\end{align}
and the instructions on the far right (namely, $\xi\rightarrow 0$ and $\mathrm{lin}\,N$) are to be imposed after taking functional derivatives. Also, we introduced the periodic function $I_{\mu}(\tau) \coloneqq \int_{0}^{\tau} \ud \tau' \big[ a_{\mu}(\tau') - \llangle a_{\mu} \rrangle \big]$ and $M^{2}(a) \equiv m^{2} -\llangle a^{2}\rrangle + \llangle a\rrangle^{2}$, which is the worldline analogy to the Kibble mass, and defined a worldline average through
\begin{equation}
    \llangle f \rrangle := \frac{1}{T} \int_{0}^{T} d\tau \, f(\tau)
\end{equation}
which can be meaningfully evaluated once the limit $\xi(\tau) \to 0$ is taken (see \cite{Copinger:2023ctz} for how the worldline averages in $M^{2}(a)$ can be converted to the \textit{space-time} averages that appear in the Kibble mass). In fact, $\mathcal{S}^{\rm tr.}_{\rm pw.}[a]$ is really the classical worldline action, which reflects the non-trivial semi-classical exactness of the worldline path integral in plane wave backgrounds as discussed for instance in \cite{AntonPlane, Edwards:2021vhg, Copinger:2023ctz}.

In \eqref{Dxx_full}, then, we have formally dealt with the non-null parameter, $\rho^{2}$, to all orders. This manifests itself, however, in two ways: the differential operators have an explicit $\rho^{2}$ dependence which we will treat further in the following sections; and an implicit dependence through the argument of $a_\mu(x_{\textrm{cl.}}^{\tri})$. The latter corresponds to a partial resummation in $\rho^{2}$, in the part that retains the functional form of the semi-classically exact action in a null plane wave background.

\subsubsection{$N$-photon dressed propagator: momentum-space representation}

The preceding discussion has outlined the techniques required to handle worldline path integrals in a non-null plane-wave background, focusing on the space-time representation of the $N$-photon dressed propagator $\mathcal{D}^{x'x}_N[a]$. For the computation of scattering amplitudes---which are ultimately the quantities of interest---it is more natural to work in momentum space. We therefore turn next to $\mathcal{D}^{p'p}_N[a]$, the momentum-space representation of the $N$-photon dressed propagator, defined as 
\begin{align}
    \mathcal{D}^{p'p}_N[a]=\int d^4x\,d^4x' \e^{ip'\cdot x-ip\cdot x}\,\mathcal{D}^{x'x}_N[a]\,.
\end{align}
A direct, brute-force approach to computing $\mathcal{D}^{p'p}_N[a]$ would be to evaluate the Fourier transform above. In a non-null background, however, performing these Fourier integrals is considerably more subtle than in the plane-wave case, and, moreover, the intermediate steps are largely superfluous. For this reason, we delegate the explicit calculation to Appendix~\ref{sec:appendix_scalar} for completeness. 

Alternatively, and more elegantly, we can instead perform the worldline path integral with a mixed-boundary condition to first compute $\mathcal{D}^{p'x}_N[a]$ so that the desired momentum space propagator can be written in the form $\mathcal{D}^{p'p}_N[a]=\int \e^{-ip\cdot x}\mathcal{D}^{p'x}_N[a]d^4x$, in direct analogy with the approach outlined in \cite{Copinger:2023ctz}. Both approaches ultimately lead to the representation
\begin{align}
    & \mathcal{D}_{N}^{p'p}[a] = i(-ie)^{N}\int d^4x \int_{0}^{\infty} dT\int\prod_{i=1}^{N}d\tau_i\,\,\,  \e^{i \rho^{2}\int \ud \tau' \ud \tau'' \tilde{\Delta}(\tau', \tau'') \frac{\delta}{\delta \xi(\tau')} \frac{\delta}{\delta \xi(\tau'')}}\e^{i\tilde{\mathcal{S}}^{\rm tr.}_{\rm pw.}[a(\bullet+\xi)]} \Big|_{\mathrm{lin}\,N}^{\xi(\tau) \rightarrow 0}\,,
\label{Dpp_full}
\end{align}
where the worldline Green's function with mixed boundary conditions is this time \cite{103}
\begin{equation}\label{eq:Delta_tilde}
    \tilde{\Delta}_{ij} \coloneqq 
    \tilde{\Delta}(\tau_{i},\tau_{j}) = \frac{1}{2}|\tau_{i}-\tau_{j}|-\frac{1}{2}(\tau_{i}+\tau_{j})\,,
\end{equation}
and the exponential is changed to (recall that the plane wave path integral is semi-classically exact)
\begin{align}
    i\tilde{\mathcal{S}}^{\rm tr.}_{\rm pw.}[a]=i\int_{0}^{T}\left(P_{N}\cdot P_{N}-m^2\right)d\tau-i(p'+K-p)\cdot x
\end{align}
with the dressed worldline momentum given by
\begin{align}
    P_{N}=\tilde{p}'-a(n\cdot \tilde{x}_{\rm cl.}(\tau))-\sum_{i = 1}^{N} [ {}^{\bullet\!}\tilde{\Delta}^{\!\bullet}(\tau, \tau_{i}) \varepsilon_{i}^{\mu} + i\,{}^{\bullet\!}\tilde{\Delta}(\tau, \tau_{i})k_{i}^{\mu}]\,
    \label{eq:P_N_1}
\end{align}
and the appropriate classical solution $\tilde{x}_{\rm cl.}(\tau)$ is given by
\begin{align}
    \tilde{x}_{\rm cl.}^{\mu}(\tau)=x^{\mu}+2p'^{\mu}\tau-\sum_{i=1}^{N}\left[|\tau-\tau_i|-(\tau+\tau_i)\right]k^{\mu}_{i}-2i\sum_{i=1}^{N}\Theta(\tau-\tau_i)\epsilon^{\mu}_{i}\,.
\end{align}
While obvious, we also emphasis that the transformation $a(\bullet+\xi)$ in this context translates to
\begin{align}
    \mathcal{F}[a(\bullet+\xi)]=  \mathcal{F}[a(\tilde{x}_{\rm cl.}^{\Delta}+\xi)]\,.
\end{align}
Interestingly, when viewed in light of the spacetime representation \eqref{Dxx_full} discussed earlier, the momentum-space expression in \eqref{Dpp_full} takes precisely the form one would naturally expect once the change in boundary conditions of the worldline path integral is properly accounted for.

\subsubsection{$N$-photon dressed effective action}

For closed trajectories, $q(\tau)$ is periodic and the expansion about the loop center of mass ($x_{0} := \llangle x \rrangle$) requires that it integrate to zero (``string inspired'' boundary conditions: SI), and we have
\begin{equation}
    \int_{SI} \hspace{-0.25em} \mathcal{D}q(\tau)\,
 \e^{i\int_{0}^{T}\ud\tau\big[-\frac{\dot{q}^{2}}{4}-\delta\mathcal{J}\cdot q\big]} = {-i(4\pi T)^{-2}}\,
 \exp\Big[
 -\frac{i}{2}\int_{0}^{T}\ud \tau_i \ud \tau_j\,  \delta\mathcal{J}_\mu(\tau_i)G_{ij} \delta\mathcal{J}^\mu(\tau_j)
 \Big] \;,
\end{equation}
where now the (bosonic) worldline Green function is 
\begin{equation}\label{eq:GB}
    G_{ij} \coloneqq G(\tau_{i},\tau_{j}) = |\tau_{i}-\tau_{j}|- \frac{(\tau_{i} - \tau_{j})^{2}}{T}-\frac{T}{6}\,.
\end{equation}
Then, repeating the same steps as for the propagator, we arrive at a similar replacement 
\begin{equation}\label{implicit_a_loop}
    a_{\mu}\big(x_{\rm cl.}^{\tri} + \xi(\tau)\big) \longrightarrow a_{\mu}(\tau) := a_{\mu}\big(x_{\rm cl.}^{\tri}(\tau)\big) \,,
\end{equation}
where this time the classical solution is given by
\begin{align}
    x_{\rm cl.}^{\mu}(\tau)=\xnot^{\mu} - i\sum_{j=1}^{N} [\Gd(\tau, \tau_{j}) \varepsilon_{j}^{\mu} - iG(\tau, \tau_{j})k_{j}^{\mu}]\,,
\end{align}
and integrating over the position of the loop centre of mass, one finds that the path integral for the scalar effective action evaluates to
\begin{align}
    &\Gamma_{N}[a] = (-ie)^{N}\int d^{4}\xnot\,\int_{0}^{\infty} \frac{dT}{T}(4\pi iT)^{-2}  \int_{0}^{T}\prod_{i=1}^{N}d\tau_i\,\e^{\frac{i}{2} \rho^{2}\int \ud \tau' \ud \tau''\, G(\tau', \tau'') \frac{\delta}{\delta \xi(\tau')} \frac{\delta}{\delta \xi(\tau'')}} \e^{i\mathcal{S}^{\rm lp.}_{\rm pw.}[a(\bullet+\xi)]} \Big|_{\textrm{lin}\,N}^{\xi(\tau) \rightarrow 0}\,.
\label{effa_full}
\end{align}
Here $\mathcal{S}^{\rm lp.}_{\rm pw.}[a]$, like in the open-line case, is the corresponding on-shell action for the null plane wave at one-loop, given explicitly by
\begin{align}
   i\mathcal{S}^{\rm lp.}_{\rm pw.}[a] &=-iM^{2}(a) T + i\sum_{j=1}^{N}  k_{j} \cdot \xnot+2\sum_{j=1}^{N} \big[\big(\llangle a \rrangle - a(\tau_{j}) \big)\cdot\varepsilon_{j}+i\big( I(\tau_{j}) - \llangle I \rrangle \big)\cdot k_{j}\big] \\\nonumber
&-\frac{i}{2} \sum_{i,j=1}^{N}[G_{ij}k_{i}\cdot k_{j} - 2i\Gd_{ij}\varepsilon_{i}\cdot k_{j} + \Gdd_{ij} \varepsilon_{i}\cdot\varepsilon_{j}]\,.
\end{align}
In fact, in direct analogy to $\mathcal{D}^{p'p}_{N}$, the above term can be written equivalently as
\begin{align}\label{eq:S_lp}
    i\mathcal{S}^{\rm lp.}_{\rm pw.}[a]=i\int_{0}^{T}\left(Q_{N}\cdot Q_{N}-m^2\right)d\tau-iK\cdot \xnot
\end{align}
where now the dressed momentum $Q_{N}$ takes the form 
\begin{align}\label{eq:def_QN}
    Q^{\mu}_{N}(\tau)&=\llangle a^{\mu}\rrangle-a^{\mu}(\tau)-\frac{1}{2}\sum_{i = 1}^{N} [ \dot{G}(\tau_{i}, \tau) k_{i}^{\mu} + i \ddot{G}(\tau_{i}, \tau) \epsilon_{i}^{\mu}]\,.
\end{align}
Note that three components of the integral over loop centre of mass can still be computed:
\begin{equation}
    \int d^{4}\xnot \, \e^{iK \cdot \xnot} = \hat{\delta}(K_{-})\hat{\delta}^{2}(\mathbf{K}_{}) \frac{1}{2}\int d\xnot^{\tri} \,\e^{iK_{\tri}\xnot^{\tri}}\,,
\end{equation}
although now $\delta(K_{-}) \sim \delta(K^{+} - \frac{\rho^{2}}{2}K^{-})$ with $K := \sum_{j=1}^{N}  k_{j}$. The last integral cannot be computed without specifying the field dependence on $x^{\tri}$. However, in the Bern-Kosower part of the exponent in \eqref{effa_full} (the double sum in the second line of \eqref{eq:S_lp}), we can further replace $G(\tau_{i}, \tau_{j}) \to G_{B}(\tau_{i}, \tau_{j}) = |\tau_{i} - \tau_{j}| - \frac{(\tau_{i} - \tau_{j})^{2}}{T}$, since the constant part drops out by momentum conservation in the $-$ and $\perp$ directions: for on-shell photons due to \eqref{eq:MassShell} imposing momentum conservation also in the $\tri$ direction, but also for \textit{off-shell} photons due to the line element \eqref{eq:ds2}. 

To summarise, a unified picture emerges for both the photon-dressed propagator and the photon-dressed effective action in a non-null plane-wave background. To make this structure explicit, we introduce the abstract notation $\mathbb{X}_{N}[a]$ to denote either the dressed propagator or the dressed effective action, so that
\begin{align}\label{eq:unified_tree_loop}
    \mathbb{X}_{N}[a]=\int \mathbb{D}_{\mathbb{X}}[...] \e^{i \rho^{2}\int \ud \tau' \ud \tau'' G_{\mathbb{X}}(\tau', \tau'') \frac{\delta}{\delta \xi(\tau')} \frac{\delta}{\delta \xi(\tau'')}} \e^{i\mathcal{S}^{\mathbb{X}}_{\rm pw.}[a(\bullet +\xi)]}\Big|_{\textrm{lin}\,N}^{\xi(\tau) \rightarrow 0}\,.
\end{align}
where 
\begin{align}
\int \mathbb{D}_{\mathbb{X}}[...]=\begin{cases}
    \int d^4 x\int_{0}^{\infty}dT\int_{0}^{T}\prod_{i=1}^{N}d\tau_i\\
    -i\int d^4 \xnot \int_{0}^{\infty}\frac{dT}{T}\int_{0}^{T}\prod_{i=1}^{N}d\tau_i
\end{cases}
\quad,\quad
G_{\mathbb{X}}=\begin{cases}
   \tilde{\Delta} \\
    \frac{G}{2}
\end{cases}
\quad\&\quad
\mathcal{S}^{\mathbb{X}}_{\rm pw.}=\begin{cases}
   \tilde{\mathcal{S}}^{\rm tr.}_{\rm pw.}\\
    \mathcal{S}^{\rm lp.}_{\rm pw.}
\end{cases}\,,
\end{align}
respectively, for $\mathbb{X}=\mathcal{D}^{p'p}$ and $\mathbb{X}= \Gamma$. Despite the compact forms, the expressions above are still formal and must be unpacked before they can be used for explicit calculations at a given order in $\rho^{2}$. In what follows, we focus on the leading-order contributions induced by the non-null background, starting with the tree-level amplitudes. From this point forward, for simplicity, we will take the gauge of the external photons to be such that $\varepsilon_j^{\tri}=0$, simplifying the implicit dependence in both the propagator and effective action and greatly facilitating the expansion to multi-linear order in these photon polarisations.

\subsection{Multi-photon tree-level scattering: LSZ, amplitudes and wavefunction}

The above analysis, while treating the non-null field exactly, does so at the expense of expressing the dressed propagator and effective action in functional derivative form. To arrive at expressions more amenable to phenomenological applications, we now turn to a resummed expansion in the non-null parameter, $\rho^2$. Led by our success in treating the gauge field's dependence on $\rho^{2}$ through its argument \textit{exactly}, our expansion scheme dictates that implicit $\rho^2$ dependence, i.e., in the background field dependence for~\eqref{implicit_a_line},~\eqref{implicit_a_loop}, be kept to all-orders, whereas the explicit dependence in $\rho^2$, such as it appears along with functional derivative terms in either the propagator or effective action, be treated perturbatively. We will apply the same approach when dealing with spin degrees of freedom below. 

We will later return to the issue of the physical interpretation of the small parameter $\rho^2$ and to the question of how it is rendered dimensionless so as to justify such an expansion; it suffices to note -- and we will demonstrate this explicitly in what follows -- that the coupling of the background and the photons to the classical worldline velocity, $\dot{x}(\tau)$, effectively reorganises the expansion parameter into the dimensionless combination $\rho^2/|\dot{x}_{\textrm{cl.}}(\tau)|$. For the time being, we assume the validity of this perturbative expansion in $\rho^{2}$ and proceed to explore its consequences for tree-level amplitudes. In practice, expressing the $N-$photon dressed propagator in this resummed scheme amounts to the following simplification 
\begin{equation}
    \e^{i \rho^{2}\int \ud \tau' \ud \tau'' \,\tilde{\Delta}(\tau', \tau'') \frac{\delta}{\delta \xi(\tau')} \frac{\delta}{\delta \xi(\tau'')}}\to 1+i \rho^{2}\int_{0}^{T} \ud \tau' \ud \tau'' \,\tilde{\Delta}(\tau', \tau'') \frac{\delta}{\delta \xi(\tau')} \frac{\delta}{\delta \xi(\tau'')} +\mathcal{O}(\rho^4)\,,
    \label{eq:func_der_small}
\end{equation}
in~\eqref{Dpp_full} and with an analogous replacement for the effective action in Eq.~\eqref{effa_full}, obtained by substituting, of course, $\tilde{\Delta}(\tau',\tau'')\to \tfrac{1}{2}G(\tau',\tau'')$. We now turn to the evaluation of scattering processes on the open line and on the loop; the former, which we address first, requires LSZ truncation and involves some additional technical subtleties.

The appropriate object $\mathcal{A}_{N}^{p'p}$ for describing tree-level scattering of scalar particles in non-null background fields, following~\cite{Copinger:2023ctz}, is defined as
\begin{equation}
    \mathcal{A}_{N}^{p'p}=\lim_{p'^2,p^2\rightarrow m^2}-(p'^2-m^2)(p^2-m^2)\mathcal{D}_{N}^{\tilde{p}'p}\,,
    \label{eq:amp_def_scal}
\end{equation}
where $\tilde{p}=p+a^\infty$ here is the kinetic momentum of the scattered particle at asymptotic infinity (we are denoting $a^{\infty} = a(x^{+} \to \infty)$ and choose $a(-\infty) = 0$). The corresponding momentum-space propagator is given in \eqref{Dpp_full}. In principle, the $\mathcal{O}(\rho^{2})$ correction to the amplitude can be obtained directly by applying the expansion \eqref{eq:func_der_small} and subsequently performing the LSZ reduction. However, it is advantageous to first simplify the linear order in $\rho^2$ expression for $\mathcal{D}^{\tilde{p}'p}$  to a convenient form before the LSZ reduction. This alternative representation, derived in Appendix~\ref{sec:appendix_scalar}, reads
\begin{align} \label{eq:DNrho}
    \mathcal{D}_{N}^{\tilde{p}'p}&=(-ie)^{N}\hat{\delta}_{-\LCperp}(p-K-\tilde{p}^{\prime})\int_{0}^{\infty}dT\,\e^{i(p'^{2}-m^{2}+i0^{+})T}\frac{1}{2}\int dx^{\tri}\, \e^{i(\tilde{p}^{\prime}+K-p)_{\tri}x^{\tri}}\prod_{i=1}^{N}\int_{0}^{T}d\tau_{i}\,\big[1+\rho^{2}\mathcal{I}_{N}+\mathcal{O}(\rho^4)\big]\\
    &\times \e^{-2iTp^{\prime}\cdot\llangle\delta a\rrangle +iT \llangle \delta a^2 \rrangle +2\sum_{i=1}^{N}[a_{i}\cdot\varepsilon_{i}+i\int_{0}^{\tau_{i}}d\tau \,a(\tau)\cdot k_{i}]+i(2\tilde{p}^{\prime}+K)\cdot g+iS_N^{\textrm{BK}}}\bigg|_{\mathrm{lin}\, N}\,,
    \notag
\end{align}
where the first order correction in the non-null parameter, $\mathcal{O}(\rho^2)$, to the propagator integrand is
\begin{equation}
    \mathcal{I}_{N}=-2i\int_{0}^{T}d\tau\int_{0}^{T}d\tau'\, |\tau-\tau'|P_{N}(\tau)\cdot a'(\tau)P_{N}(\tau')\cdot a'(\tau')\,,
    \label{eq:I_N}
\end{equation}
with 
$P_{N}(\tau)\to\tilde{p}'-a(\tau)+\sum_{i=1}^{N}[\Theta(\tau_{i}-\tau)k_{i}-\delta(\tau_{i}-\tau)i\varepsilon_{i}]$ corresponding to an effective Green function coinciding with that of the inifinite line; c.f., ~\eqref{eq:P_N_1}. The Bern-Kosower portion of the action of the momentum space propagator now reads
\begin{equation}
    S_N^{\textrm{BK}}\coloneqq -\sum_{i,j=1}^{N}\Bigl\{\frac{1}{2}|\tau_{i}-\tau_{j}|k_{i}\cdot k_{j}-i\mathrm{sgn}(\tau_{i}-\tau_{j})\varepsilon_{i}\cdot k_{j}+\delta(\tau_{i}-\tau_{j})\varepsilon_{i}\cdot\varepsilon_{j}\Bigr\}\,.
\end{equation}
Lastly, the implicit field dependence (containing the exact resummed non-null dependence) now becomes
\begin{equation}
    a_{i}^{\mu}=a^{\mu}\Bigl(x^{\tri}+g^{\tri}+(\tilde{p}^{\prime\tri}+p^{\tri})\tau_{i}-\sum_{j=1}^{N}k_{j}^{\tri}|\tau_{i}-\tau_{j}|\Bigr)\,.
    \label{eq:implicit}
\end{equation}
Surprisingly, despite the boundary conditions for the line manifestly breaking the translational invariance of the worldline theory, the $\mathcal{O}(\rho^{2})$ contribution is written here in terms of functions depending only on the relative separation of the insertion points of external photons or the coupling to the background (this is not the case at $\mathcal{O}(1)$ in our expansion until \textit{after} the LSZ prescription draws the external matter legs to asymptotic infinity).

Having recast the propagator into this more convenient form, we are now in a position to carry out the LSZ truncation and thereby obtain the amplitude \eqref{eq:amp_def_scal}. In background fields that asymptotically approach constant values, it was found that truncating matter lines amounts to a projection of the Schwinger propertime integral onto $T\to \infty$ and $T\to -\infty$ for outgoing state wavefunction and incoming state wavefunctions respectively~\cite{Copinger:2023ctz,Copinger:2024twl}. We now illustrate how this reduction proceeds in the present case of non-null background fields.

We begin with the truncation of the outgoing leg. Using the relation and standard Feynman regulation~\cite{Copinger:2023ctz}
\begin{equation}
    \lim_{p'^2\rightarrow m^2-i0^\LCp}-i({p'}^2-m^2+i 0^\LCp) \int_0^\infty\!d T\, \e^{ i ({p'}^2-m^2 + i 0^\LCp) T} F(T)=\lim_{p'^2\rightarrow m^2}F(\infty)\,,
    \label{eq:FT}
\end{equation}
where $F(T)$ represents the integrand of the propagator (e.g. in \eqref{eq:DNrho}, but the relation is valid to all orders in $\rho^{2}$) excluding the free part of the action, we can see that truncation on the outgoing left projects the propertime argument of the integrand in the propagator expression onto the limit $T\to\infty$. While the object thus obtained -- as a result of applying the LSZ truncation (once) to the outgoing leg -- serves here as an intermediate step in the computation of the $\mathcal{A}^{p'p}_{N}$, it can also be thought of as the $N$-photon-dressed generalisation of the momentum-space wavefunction. In fact, in Appendix \ref{app:wavefn_compton} we explicitly verify that in the case $N=0$, this intermediate object reproduces the known wavefunction in the non-null plane wave background~\cite{Heinzl:2016kzb,Mackenroth:2018rtp,Mackenroth:2020pct}.

We next turn to the truncation of the incoming leg. Following the method introduced in ~\cite{Mogull:2020sak}, we define a collective proper-time coordinate
\begin{equation}
    \tau_0 \coloneqq \frac{1}{N}\sum_{i=1}^N\tau_i \;,
    \qquad
    \bar{\tau}_i\coloneqq \tau_i-\tau_0\,,
\end{equation}
which satisfies the constraint $\sum_{i=1}^{N}\bar{\tau}_i=0$. As a result, only $N-1$ of the $\bar{\tau}_i$ are independent, with the constraint enforced later via a delta function. Completing the change of variables to $\tau_0$ and the $N-1$ independent $\bar{\tau}_i$, the Jacobian yields a factor of $N$, and the integration ranges are constrained by $\tau_0+\bar{\tau}_i\geq 0$ and $\tau_0-\sum_{i=1}^{N-1}\bar{\tau}_i\geq 0$. This leads to
\begin{equation}
    \prod_{i=1}^N\int_{0}^{\infty}\!\ud\tau_i\to N\int \ud \tau_0
    \prod_{i=1}^{N-1}\ud \bar{\tau}_i \prod_j^\infty\Theta(\tau_0+\bar{\tau}_j)\Theta\Bigl(\tau_0-\sum_{i=1}^{N-1}\bar{\tau}_i\Bigr)\,.
\end{equation}
Introducing the final coordinate $\bar{\tau}_N$ via a delta function $\delta(\sum_{j=1}^N\bar{\tau}_j)$, the external proper-time integrations may be written in the compact form
\begin{equation}
    \prod_{i=1}^N\int_{0}^{\infty}\!\ud\tau_i
    =\int_{0}^{\infty}\!\ud\tau_0
    \prod_{i=1}^N\int_{-\tau_0}^{\infty}\!\ud \bar{\tau}_i \, \delta\bigg(\sum_{j=1}^N\frac{\bar{\tau}_j}{N}\bigg) \,.
    \label{eqAmpp}
\end{equation}
The redefinition supplies us with an additional global Schwinger propertime-like integral which performs (in an analogous way to the outgoing state) truncation on the incoming state, extending the range of integration to the entire real line. We facilitate this step through the redefinition $\bar{x}^{\tri}:=x^{\tri}+(p'+p+K)^{\tri}\tau_{0}+g^{\tri}(\{\bar{\tau}_{i}\})$ with \linebreak $g \equiv g(\{\tau_i\})\coloneqq \sum_{i=1}^{N}(k_i\tau_i-i\varepsilon_i)$. After rearranging terms and applying ~\eqref{eq:FT} to the $\tau_0$ integral -- which now projects $\tau_0\to\infty$ and extends the integration range from $\int_0^\infty d\tau$ to $\int_{-\infty}^{\infty} d\tau$ -- we arrive at the final expression for the scalar amplitude dressed with $N$ photons,
\begin{align}
    \mathcal{A}_{N}^{p'p}&=(-ie)^{N}\int d^{4}x\, \e^{i(K+\tilde{p}'-p)\cdot x}\int_{-\infty}^{\infty}\prod_{i=1}^{N}d\tau_{i}\,\delta\bigg(\sum_{j=1}^{N}\frac{\tau_{j}}{N}\bigg)\big[1+\rho^{2}\mathcal{I}_{N}\big] \e^{i({\tilde p}' +p)\cdot g+iS_N^{\textrm{BK}}}\notag\\
    &\times \e^{-i\int_{-\infty}^{0}d\tau\, [2\tilde{p}'\cdot a(\tau)-a^{2}(\tau)]-i\int_{0}^{\infty} d\tau\, [2p'\cdot\delta a(\tau)-\delta a^{2}(\tau)]-2i\sum_{i=1}^{N}[\int_{-\infty}^{\tau_{i}}d\tau \, k_{i}\cdot a(\tau)-i\varepsilon_{i}\cdot a(\tau_{i})]}\Big|_{\mathrm{lin}\, N} \;,
    \label{eq:scalar_amp_final}
\end{align}
where now the explicit $\rho^2$ dependent correction, $\mathcal{I}_N$, is given in~\eqref{eq:I_N} but with $\int^T_0 d\tau\to\int^\infty_{-\infty}d\tau$. Also, the implicit dependence finally reads
\begin{equation}
    a^\mu(\tau)=a^\mu\Bigl(x^{\tri}+(p^{\prime\tri}+p^{\tri})\tau-\sum_{i=1}^{N}k_{i}^{\tri}|\tau-\tau_{i}|\Bigr)\,,
    \label{eq:implicit_LSZ}
\end{equation}
where, for simplicity, we have relabeled $\bar{\tau}\to\tau$. Note that, as in the plane wave case, the momentum conservation in the ``minus'' and transverse directions, $(K+\tilde{p}'-p)_{\LCm\LCperp}=0$, remains manifest.

In this compact worldline representation, the amplitude encodes tree-level scattering with an arbitrary number $N$ of external photons. One may determine non-null deviations at arbitrary order in $\rho^{2}$ away from the idealistic plane wave profile useful for phenomenological studies as are relevant for current and upcoming high-powered laser facilities. One may also easily make the connection to the plane wave case~\cite{Copinger:2023ctz} by setting $\rho^2\to 0$. As a further non-trivial check of our result, we have explicitly verified our formula for $N=1$ against the non-linear Compton amplitude computed from the standard Feynman diagrammatic approach (see Appendix \ref{app:crosscheck}). We emphasise that this exercise of verification -- which involves a chain of non-trivial algebraic tricks to recast our worldline parameter integrals into the form of integrals over the field's phase -- more than serves the purpose of a cross-check; it demonstrates that the worldline method that we adopted is significantly more efficient at arriving at a convenient, ready-to-use expression for the amplitude. 

While this concludes our main results for the case of open worldlines, we now briefly return to the question of the formal expansion parameter, before moving to the discussion at one loop order.

\subsubsection{$\rho^2$ as an expansion parameter}\label{sec:rho_exp_tree}
\label{sec:ExpLine}
Going through the conversion from the propertime representation to spacetime coordinates, while cumbersome, serves the dual purpose of providing a check on the wavefunction and non-linear Compton scattering, and furnishing us with a reliable expansion parameter. Since $\rho^2$ actually arises from $\omega^2$, which carries with it a dimension of momentum squared, we must carefully identify the roots of different dimensionless factors that appear in the calculations. For instance, in Appendix \ref{app:crosscheck}, we show that doing a change of variables from propertime to spacetime, the (non-null) wavefront phase direction is extracted, giving us the expansion parameter is proportional to $\rho^2/(2p^{\prime \tri})^2$ in the case of the wavefunction, \eqref{eq:scalar_wavefunction_final}---the correct dimensionless factor would be $\omega^2/(2p^{\prime \tri})^2$. Likewise, for non-linear Compton scattering, both the combinations $\rho^2/(2p^{\prime \tri})^2$ and $\rho^2/(2p^{\tri})^2$ (conservation of momentum), were found to be arise. And indeed in~\cite{Heinzl:2016kzb} a similar parameter was found to be the suitable expansion parameter for the resummed wavefunction after a slowly varying envelope approximation~\cite{PhysRevE.83.026406} (see also~\cite{raicher2015novel,Raicher:2016bbx}).

In addition to furnishing compact formulae, the worldline form for the scattering amplitude further provides a succinct criterion for the expansion parameter for \textit{any} number of external photons. Whether explicit, such as is written for the $\mathcal{I}_N$,~\eqref{eq:I_N}, or implicit in the background field definition,~\eqref{eq:implicit_LSZ}, to extract the spacetime dependent properties one need only compare to the classical $\dot{x}^{\tri}_{\rm cl.}$. Let's see how this is accomplished for $\mathcal{I}_N$. We would like to change variables from $\tau$ to
\begin{equation}
    \varphi = x_{\rm cl.}^{\tri}(\tau)=x^{\tri}+(p^{\prime\tri}+p^{\tri})\tau-\sum_{i=1}^{N}k_{i}^{\tri}|\tau-\tau_{i}|\,,
    \label{eq:change_var}
\end{equation}
and to do so let us break up the propertime integral into ordered segments, i.e.,
\begin{equation}
    \int^\infty_{-\infty}d\tau \to \prod_{i,j=1;\,i\neq j}^N[\Theta(\tau_i-\tau_j)+\Theta(\tau_j-\tau_i)]\int^\infty_{-\infty}d\tau\, \prod_{i=1}^N[\Theta(\tau-\tau_i)+\Theta(\tau_i-\tau)]\,.
    \label{eq:breakup}
\end{equation}
And then for each $N^3$ segment we can see that there exists an inverse such that $\tau(x_{\rm cl.})$ may be found, e.g. \linebreak $\tau = (x_{\rm cl.}^{\tri}(\tau)-x^{\tri})/(p^{\prime\tri}+p^{\tri})$ for $N=0$. The key point is that one may accomplish the variable transform to the lightfront spacetime coordinate by breaking up the integral, and in so doing, after re-assembling, an expansion parameter may be determined. Then one may re-express $\mathcal{I}_N$ after re-assembly as 
\be
    \rho^2\mathcal{I}_{N}=-2i\int_{-\infty}^{\infty}d\varphi\left( \frac{\rho^2}{\dot{\varphi}}\right)\int_{-\infty}^{\infty}d\varphi'\left(\frac{1}{\dot{\varphi}'}\right)|\tau(\varphi)-\tau(\varphi')|P_{N}(\varphi)\cdot a'(\varphi)P_{N}(\varphi')\cdot a'(\varphi')\,,
\ee
where the integrals over $\varphi$ are understood to be defined piecewise over the segments in which $\tau(\varphi)$ is well defined. To express this factor in a more covariant way, we can now reinstate the phase-momentum $\omega=\Omega\, \nn$ of the background field, as we introduced it in Section~\ref{sect:non-null} and state the condition for the perturbative expansion to be reliable as:
\be
    \frac{\omega^2}{|\omega\cdot \dot{x}_{\rm cl.}(\varphi)|}\ll 1\,,
\ee
which provides a concrete realisation of the condition in \eqref{eq:small_rho_2}. An equivalent way of expressing this requirement is:
\be
    \frac{\Omega\rho^2}{|\dot{x}_{\rm cl.}^{\tri}(\tau)|}=\frac{\Omega\rho^2}{|\dot{x}_{\rm cl.}^{+}(\tau)|}+\mathcal{O}(\rho^4)\ll 1\,,
    \label{eq:exp_para}
\ee
with the understanding that this is evaualted at points $\tau$ away from where $\dot{x}_{\rm cl.}^{\tri}(\tau)$ is discontinuous.

In summary, the small $\rho^2$ expansion for the tree-level amplitudes appears to correspond to a kind of large transverse momentum expansion. One may then ask whether satisfying the above condition imposes a restriction on the kinematics.
This restriction can be stated as\footnote{We remark this is quite similar to the case of the master formula for $N-$scattered photons of impulsive PP-wave to be identified with the amplitude of the free $N+1$ case in~\cite{Copinger:2024twl}, in which a ``positivity constraint" (or $\dot{x}_{\rm cl.}^{\LCp}(\tau)> 0$ or $p^{\prime \LCp}, p^{\prime \LCp}+\sum_{i\in \mathcal{U}}k_i^\LCp>0\;\forall\,\mathcal{U}\subseteq \{1,2,...,N\}$) was used.} $|\dot{x}_{\rm cl.}^{\LCp}(\tau)| \gg1/(\Omega\rho^2)$ or $p^{\prime \LCp}, p^{\prime \LCp}+\sum_{i\in \mathcal{U}}k_i^\LCp\gg1/(\Omega\rho^2)\;\forall\,\mathcal{U}\subseteq \{1,2,...,N\}$.
To gain some intuition for this requirement, note that when the incoming transverse momentum is large (i.e., $p^{\LCp}\gg1/(\Omega\rho^2)$) the above kinematic condition is automatically satisfied in the cases where all photons are either incoming or outgoing. The full set of kinematic configurations satisfying this bound is, of course, broader than these simple examples.

\subsection{Multi-photon one-loop scattering}
\label{sec:OneLoop}
For the one-loop amplitudes, the procedure is similar, although we no longer have the complications of LSZ reduction (and in the worldline formalism, external photon legs are already properly amputated). We again apply \eqref{eq:func_der_small} to arrive at the analogous equation to \eqref{eq:DNrho}
\begin{align}
    \Gamma_{N}[a] = \frac{1}{2}(-ie)^{N}\hat{\delta}(K_{-})\hat{\delta}^{2}(\mathbf{K}_{\perp}) \int_{0}^{\infty} \frac{dT}{T} (4\pi i T)^{-2}\int_{-\infty}^{\infty} d\xnot^{\tri} \,\e^{iK_{\tri}\xnot^{\tri}}\,\prod_{i=1}^{N}\int_{0}^{T}d\tau_{i} \, \e^{-i M^{2}(a) T} \big[ 1 + \rho^{2} \mathcal{J}_{N}\big] \nonumber\\
    \times \e^{2\sum_{j=1}^{N} \big[\big(\llangle a \rrangle - a(\tau_{j}) \big)\cdot\varepsilon_{j}+i\big( I(\tau_{j}) - \llangle I \rrangle \big)\cdot k_{j}\big] 
-\frac{i}{2} \sum_{i,j=1}^{N}[G_{ij}k_{i}\cdot k_{j} - 2i\Gd_{ij}\varepsilon_{i}\cdot k_{j} + \Gdd_{ij} \varepsilon_{i}\cdot\varepsilon_{j}]
} \Big\rbrace \Big|_{\textrm{lin}\,N}\,,
\end{align}
where now 
\begin{align}
    \mathcal{J}_{N} = \frac{i}{2}\int_{0}^{T}d\tau' \int_{0}^{T}d\tau'' \, G(\tau', \tau'') \, \Big[ -&4 Q_N(\tau') \cdot a'(\tau') Q_N(\tau'') \cdot a'(\tau'') -2i Q_N(\tau')\cdot a''(\tau'') \delta(\tau' - \tau'') \nonumber\\
    +&i a'(\tau')\cdot a'(\tau'')\ddot{G}(\tau', \tau'')   \Big]\,,
    \label{eq:JN}
\end{align}
with $Q_{N}$ defined as in \eqref{eq:def_QN} and $a_{\mu}(\tau)$ given in \eqref{implicit_a_loop}. Note that, unlike for the open line, the functional derivative (with respect to $\xi(\tau)$) acting on $Q(\tau)$ has produced a non-zero contribution, recorded in the second and final terms in square brackets of \eqref{eq:JN}.

Let us now specialise to the effective action, whereby $N = 0$, and consider the renormalisation of the theory in the non-null background. Fixing $N = 0$ the classical solution, $x_{\rm cl.}(\tau) = x \equiv \llangle x \rrangle$, is just the constant loop centre of mass. In this case, then, $\llangle a(\tau) \rrangle = a(x^{\tri})$, so that $M^{2}(a) = m^{2}$ and $Q_{0} = 0$. With this
\begin{equation}
    \mathcal{J}_{0} = -\frac{1}{2} a'(x^{\tri})^{2} \int_{0}^{T}d\tau' \, \int_{0}^{T}d\tau'' G(\tau', \tau'') \ddot{G}(\tau', \tau'') = \frac{T^{2}}{6} a'(x^{\tri})^{2}\,.
    \label{eq:J0}
\end{equation}
The bare effective action contains a standard UV divergence produced from the $T \rightarrow 0$ region of the proper time integral. This divergence splits into a field-independent part which we remove by subtracting $\Gamma_{0}[0]$ (corresponding to normal ordering). This equates with the renormalisation in the null field case and cancels the $\mathcal{O}(1)$ part of the effective action (this agrees with the fact that the renormalised effective action in a plane wave vanishes). We are left, then, with
\begin{align}
    \Gamma_{R}[a] := \Gamma_{0}[a] - \Gamma_{0}[0] &= \frac{\rho^{2}}{12} V^{3} \int_{0}^{\infty} dT \frac{T^{1-\frac{D}{2}}}{(4\pi i)^{\frac{D}{2}}} \e^{-im^{2}T} \int_{-\infty}^{\infty} dx^{\tri}\, a'(x^{\tri})^{2} + \mathcal{O}(\rho^{4})\\
    &= \frac{\rho^{2}}{12} V^{3} (m^{2})^{\frac{d}{2} - 2} \Gamma\Big[2-\frac{d}{2}\Big] \int_{-\infty}^{\infty} dx^{\tri}\, a'(x^{\tri})^{2} + \mathcal{O}(\rho^{4})
    \label{eq:GammaR}
\end{align}
where $V^{3} = \hat{\delta}^{3}(0)$ and we have used dimensional regularisation to control a UV divergence. This divergence is clearly associated with (on-shell) charge renormalisation once we recall that $2\rho^{2}a_0^2=-\rho^2a'^2=-\frac{1}{2}\rf_{\mu\nu}\rf^{\mu\nu}$ (see \eqref{eq:F_1_def}) so we are obtaining something proportional to the Maxwell term in the bare action.

\subsubsection{Non-null CCF meets Euler-Heisenberg}
\label{sec:nnCCF}

In this section, we revisit what is probably the most well-understood problem in strong-field QED -- the Euler-Heisenberg effective action -- through the lens of the non-null field framework introduced in this work. We do this in two ways: we show that the effective action for a constant crossed field provides an example in which the $\rho^2$ expansion of \eqref{effa_full} can be formally resummed, reproducing the familiar one-loop Euler-Heisenberg result~\cite{Heisenberg:1935qt,Weisskopf:1936hya}, and then we consider the $\mathcal{O}(\rho^{2})$ contribution from our Master Formula and how it can be obtained from the original worldline path integral in our non-null field.

We start by taking $N=0$ in \eqref{effa_full}, leading to
\begin{align}
    \Gamma[a] = \int d^{4}\xnot\,\int_{0}^{\infty} \frac{dT}{T}(4\pi iT)^{-2}  \,\e^{\frac{i}{2} \rho^{2}\int \ud \tau' \ud \tau'' \, G(\tau', \tau'') \frac{\delta}{\delta \xi(\tau')} \frac{\delta}{\delta \xi(\tau'')}} \e^{i\mathcal{S}^{\rm lp.}[a(\bullet+\xi)]} \Big|_{\xi(\tau) \rightarrow 0}\,,
\end{align}
where
\begin{align}
    \mathcal{S}^{\rm lp.}[a(\bullet+\xi)]=\int_{0}^{T}d\tau \left(\braket{a^2(x^{\Delta}+\xi)}-\braket{a(x^{\Delta}+\xi)}^2-m^2\right) 
\end{align}
is the appropriate worldline action in the $N=0$ limit. So with $a_{\mu}(x^{\tri}) = a_{0}\delta_{\mu\perp} x^{\tri}$ the effective action simplifies to
\begin{align}\label{eq:Gamma_as_det}
    \Gamma[a] = \int d^{4}\xnot\,\int_{0}^{\infty} \frac{dT}{T}(4\pi iT)^{-2}  \,\e^{\frac{i}{2} \rho^{2}\int_{0}^{T} \ud \tau' \int_{0}^{T} \ud \tau'' G(\tau', \tau'') \frac{\delta}{\delta \xi(\tau')} \frac{\delta}{\delta \xi(\tau'')}} \e^{-im^2T-i2a_{0}^{2}\int_{0}^{T}d\tau\int_{0}^{T}d\tau'\,\xi(\tau)\delta(\tau-\tau')\xi(\tau')}\,.
\end{align}
Then one may formally complete the functional derivative to determine that\footnote{We used the identity \(\e^{\frac{1}{2}A_{ij}\partial_{i}\partial_j}\left(\e^{-x^{T}B \,x}\right)\Big\rvert_{x\rightarrow 0}=\big[\det\left(1+2AB\right)\big]^{-1/2}\). See Appendix~\ref{app:quadratic_operator_gaussian} for details. }
\begin{equation}
    \Gamma[a] = \int d^{4}\xnot\,\int_{0}^{\infty} \frac{dT}{T}(4\pi iT)^{-2} \e^{-im^2T}\Bigl[\mathrm{Det}\Bigl(1-4\rho^2 a_0^2 \hat{G} \Bigr) \Bigr]^{-1/2}\,.
\end{equation}
In fact, expanding the determinant, using $\det(\hat{O})=\exp({\rm Tr}(\log\hat{O}))$, reveals explicitly how an infinite set of one-loop worldline diagrams is resummed into the compact expression above.
\begin{align}
    \Bigl[\mathrm{det}\Bigl(1-4\rho^2 a_0^2 \hat{G} \Bigr) \Bigr]^{-1/2}&=\exp\left[\frac{1}{2}\sum_{l=1}^{\infty}\frac{1}{l}(4\rho^2a_0^2)^l\,{\rm Tr}(\hat{G}^l)\right]\,,\\
    &=\exp\left(\frac{1}{2}\times
  \begin{tikzpicture}[baseline]
      \draw (0,0) circle (.25cm); 
      \draw[fill=blue] (0,.25)  circle (1pt);
  \end{tikzpicture}  + 
\frac{1}{4}\times \begin{tikzpicture}[baseline]
      \draw (0,0) circle (.25cm); 
      \draw[fill=blue] (0,.25) circle (1pt); 
      \draw[fill=blue] (0,-0.25) circle (1pt); 
  \end{tikzpicture}  +
  \frac{1}{6}\times \begin{tikzpicture}[baseline]
      \draw (0,0) circle (.25cm); 
      \draw[fill=blue] (.25,0) circle (1pt); 
      \draw[fill=blue] (-0.125, 0.216506) circle (1pt); 
      \draw[fill=blue] (-0.125, -0.216506) circle (1pt); 
  \end{tikzpicture}  
   +... \right)\,,
\end{align}
where we introduce the blue-dot vertex to represent the perturbative non-null correction factor $4\rho^2a_0^2$ and the lines represent worldline propagators, $G$. To perform the resummation, we first note that 
\begin{align}
    {\rm Tr}(\hat{G}^l)&=2^l\sum_{s=1}^{\infty}\left(\frac{iT}{2\pi s}\right)^{2l}=2\left(\frac{-T^2}{2\pi^2}\right)^{l}\zeta(2l)\,,\\
    &=-\frac{T^{2l}\,2^{l+1}B_{2l}}{(2l)!}\,,
\end{align}
where the Riemann $\zeta$-function is given $\zeta(l)\equiv \sum_{s=1}^{\infty}s^{-l}$, and $B_{n}$ are the Bernoulli numbers. Now, these terms can be summed using the identity derived from their generating function (see 4.19.9 of \cite{NIST:DLMF})
\begin{align}
    -\sum_{l=1}^{\infty}\frac{2^{2l}B_{2l}\,x^{2l}}{2l (2l)!}=\log\left(\frac{x}{\sinh x}\right),
\end{align}
to arrive at the effective action in a non-null CCF background, given by
\begin{equation}
    \Gamma[a] = \int d^{4}\xnot\,\int_{0}^{\infty} \frac{dT}{T}(4\pi iT)^{-2} \e^{-im^2T}\frac{\sqrt{2\rho^{2} } T a_{0}}{\sinh \big(\sqrt{2\rho^{2}} T a_{0}\big)}\,.
    \label{eq:GammaScal}
\end{equation}
One can obtain the same result from the crossed field limit of the Euler-Heisenberg action~\cite{Heisenberg:1935qt,Weisskopf:1936hya}, prior to renormalisation. On the worldline, this is achieved by absorbing the constant background into the worldline kinetic term so that the worldline path integral in the constant non-null fields considered here takes the form
\begin{align}
    &\int d^{4}x \int_{SI} \mathcal{D}x(\tau) \e^{-\frac{i}{4} \int_{0}^{T} d\tau \, x^{\mu} \big[ -\eta_{\mu\nu} \partial_{\tau}^{2} + 2 f_{\mu\nu} \partial_{\tau}\big]x^{\nu} }\\
    =-&  \int d^{4}x\,\textrm{Det}'{}^{-\frac{1}{2}} \Big[\eta - 2f \partial_{\tau}^{-1} \Big] \\
    =-&\int d^{4}x\,\textrm{det}{}^{-\frac{1}{2}} \Big[\frac{\sinh(efT)}{efT} \Big]\,,
\end{align}
where the $\prime$ indicates that the determinant is taken on the space of functions orthogonal to the normal mode present for PBC). The matrix whose determinant we seek can be written to all orders in $\rho$ by using the formula (\ref{eq:feven}), and yields
\begin{equation}
   \textrm{det}{}^{-\frac{1}{2}} \Big[\frac{\sinh(efT)}{efT} \Big] =-\det{}^{-\frac{1}{2}} \Big[\eta + \frac{1}{\rho^{2}T^{2} a_{\perp}^{2}  }\Big(\frac{\sinh\big( T\sqrt{ \rho^{2} a_{\perp}^{2}} \big)}{  T\sqrt{\rho^{2} a_{\perp}^{2}}} -1\Big)f^{2} \Big]
\end{equation}
The determinant can be evaluated in closed form to reproduce the integrand in \eqref{eq:GammaScal}.

Expanding the effective action in the combination $\rho^{2} a_{\perp}^{2}$, one encounters ultraviolet divergences, which can be treated in complete analogy with the constant electromagnetic background case. At $\mathcal{O}(1)$ in the non-null correction, the divergence corresponds to the usual vacuum energy contribution. 
At $\mathcal{O}(\rho^2)$ one finds
\begin{align}
    \Gamma[a] &= \int d^{4}\xnot\,\int_{0}^{\infty} \frac{dT}{T}(4\pi iT)^{-2} \e^{-im^2T}\left[1+\frac{1}{2}\times \,
  \begin{tikzpicture}[baseline]
      \draw (0,0) circle (.25cm); 
      \draw[fill=blue] (0,.25)  circle (1pt);
  \end{tikzpicture}\, +\mathcal{O}(\rho^4)\right],\\
  &=\int d^{4}\xnot\,\int_{0}^{\infty} \frac{dT}{T}(4\pi iT)^{-2} \e^{-im^2T}\left[1 -\frac{1}{3}\rho^2a_0^2T^2+\mathcal{O}(\rho^4)\right]\,,
\end{align}
since ${\rm Tr}(\hat{G})=-T^2/6$. The $\mathcal{O}(\rho^2)$ term is likewise divergent; however, it is proportional to the Maxwell action (again, $2\rho^{2}a_0^2=-\rho^2a'^2=-\frac{1}{2}\rf_{\mu\nu}\rf^{\mu\nu}$) and agrees with the charge renormalisation in \eqref{eq:GammaR} when restricted to the non-null CCF. It can therefore be absorbed into a renormalisation of the classical electromagnetic term, yielding the renormalised effective action,
\begin{align}
    \Gamma_{R}[a] &= \int d^{4}\xnot\,\int_{0}^{\infty} \frac{dT}{T}(4\pi iT)^{-2} \e^{-im^2T}\bigg[\frac{\sqrt{2\rho^{2} } T a_{0}}{\sinh \big(\sqrt{2\rho^{2}} T a_{0}\big)}-1+\frac{1}{3}\rho^2a_0^2T^2\bigg]\,.
\end{align}
The effective action obtained above smoothly interpolates between the purely electric and purely magnetic limits for $-4 \leqslant \rho^{2} \leqslant 4$, as is evident from the field configuration given in \eqref{eq:EBCrossed}. 
This interpolation acquires a clear physical interpretation when one examines the stability of the vacuum via the pair-creation rate, which is determined by the imaginary part of the effective action. Performing the rotation $T \e^{-i0^+} \to -iT$ converts the oscillatory phase $\e^{-im^2T \e^{-i0^+}}$ into the exponentially damped factor $\e^{-m^2T}$, allowing the real and imaginary parts of the effective action to be identified from the analytic structure of the integrand. If poles lie on the new Euclidean contour, they generate a non-vanishing imaginary contribution, signaling vacuum instability.

In the purely electric limit, $\rho^{2} = 4$, such poles yield the standard non-perturbative pair-production rate. More generally, for $0 < \rho^{2} < 4$, the electric field dominates (as discussed at the end of Section \ref{sec:coords}) and the poles remain on the imaginary proper-time axis, leading again to vacuum instability and a non-vanishing pair-creation probability. Wick rotating and summing over (half) the residues of these poles leads to Ritus-like result for our non-null field:
\begin{equation}
    \Im \mathcal{L}^{(1)} = - \frac{2\rho^{2}a_{0}^{2}}{16\pi^{3}} \sum_{n = 1}^{\infty} \frac{(-1)^{n}}{n^{2}}\, \e^{- \frac{n \pi m^{2}}{\sqrt{2\rho^{2}} a_{0}}}\,.
\end{equation}

In contrast, when $-4 \leqslant \rho^{2} < 0$, the magnetic field dominates (again see end of Section \ref{sec:coords}). The poles are then shifted onto the real axis, implying that the imaginary part of the effective action is zero, so that the vacuum remains stable and pair production is again absent. Finally, as $\rho \to 0$ one approaches the null (plane-wave) limit of constant crossed fields with equal electric and magnetic magnitudes. In this case, the effective action reduces to a constant contribution corresponding solely to vacuum energy renormalisation, and no pair creation occurs.

\subsubsection{$\rho^2$ as an expansion parameter}\label{sec:rho_exp_loop}

Here we demonstrate, as anticipated from our discussion of the open line in Section \ref{sec:ExpLine}, that also on the loop we may use the ratio of the non-null parameter to the classical velocity as an expansion parameter. To do this, we make the following change of variables:
\begin{equation}
    \varphi = x_{\rm cl.}^{\tri}(\tau)=
    \xnot^{\tri} - \sum_{j=1}^{N}  G(\tau, \tau_{j})k_{j}^{\tri}\,,
\end{equation}
where we have applied the gauge choice of $\varepsilon^{\tri}_j=0$. Then, applying the above change of variables to the explicit $\rho^2$ term in the photon amplitude, we find that
\pagebreak
\begin{align}\label{eq:small_condtn_loop}
    \rho^2\mathcal{J}_{N} = \frac{1}{2}\oint d\varphi' \oint d\varphi'' \, \frac{\rho^2}{\dot{\varphi}'\dot{\varphi}''}G(\tau'(\varphi'), \tau''(\varphi'')) \, \Big[&  Q(\varphi') \cdot a'(\varphi') Q(\varphi'') \cdot a'(\varphi'') +  Q(\varphi')\cdot a''(\varphi'') \delta(\tau'(\varphi') - \tau''(\varphi'')) \nonumber\\
    -&\frac{i}{T} a'(\varphi')\cdot a'(\varphi'')\ddot{G}(\tau'(\varphi'), \tau''(\varphi''))   \Big]\,,
\end{align}
where we now have that
\begin{equation}
    \dot{\varphi}=\dot{x}_{\rm cl.}^{\tri}(\tau)= -\sum_{j=1}^{N}  \dot{G}(\tau, \tau_{j})k_{j}^{\tri}=-\sum_{j=1}^N\big[\mathrm{sgn}(\tau-\tau_j)-\frac{2}{T}(\tau-\tau_j)\big]k_j^{\tri}\,.
\end{equation}
Following a similar argument as we did for the tree-level scattering case, we find the same expansion parameter of~\eqref{eq:exp_para}, $\Omega\rho^2/|\dot{x}_{\rm cl.}^{\tri}(\tau)|\ll 1$, for all $\tau$ and $\{\tau'_{j}\}$ away from $\tau=\tau_i$. However, in contrast to the tree-level case -- where this condition is related to a clear restriction on the space-time kinematics -- it does not translate into a concrete kinematic constraint at loop level. On the other hand, the condition in \eqref{eq:small_rho_1} can still be satisfied, provided $\rho^2$ is sufficiently small such that the term in \eqref{eq:small_condtn_loop} is perturbative.

This discussion completes our treatment of scalar QED in this class of non-null backgrounds. There remains much that could be done, such as extending the calculation of the wavefunction (tree level, $N = 0$) or non-linear Compton amplitude ($N = 1$) to higher order in $\rho^{2}$, or looking at the double non-linear Compton amplitude ($N = 2$). On the loop, one could consider higher order corrections to the effective action ($N = 0$) in a generic background or, for example, consider the vacuum polarisation ($N = 2$). We leave this for future work and in the following sections, we move on to consider the same processes in spinor QED.

\section{Spinor QED in Non-Null Backgrounds}\label{sect:spinor}

The spinor QED version of the preceding discussion proceeds in a largely analogous manner, with the new feature being the coupling of spin degrees of freedom to the background field (and, of course, the external photons as usual). We incorporate these effects within the worldline formalism by introducing Grassmann variables. In particular, one finds that the worldline path integrals in non-null backgrounds again give rise to exact formal relations analogous to \eqref{eq:unified_tree_loop}, for appropriately defined quantities. However, explicitly deriving the spinor analogue of \eqref{eq:unified_tree_loop} would largely amount to a repetition of the steps already carried out in the scalar case. Accordingly, in this section, we focus on highlighting the spin-dependent technical aspects, rather than reproducing the direct analogue of \eqref{eq:unified_tree_loop} step by step. For concreteness, we illustrate these points explicitly using the leading-order non-null corrections to both one-loop and tree-level quantities in spinor QED.

\subsection{N-photon dressed propagator and effective action}

Let us first record the worldline path integral representation of the matter propagator in spinor QED dressed with N-photons. For an arbitrary background, the worldline approach appeals to the \textit{second order} formalism of the Dirac field by using the Gordon identity to arrive at the propagator in the form \cite{Ahmadiniaz:2020wlm}
\be
     \mathcal{S}^{x'x}[A] = \big(-i\slashed{D}_{x'} - m\big)\mathcal{K}^{x'x}[A]\;,
\ee
whose kernel, $\mathcal{K}^{x'x}[A]$, is given by the worldline path integral
\be
    \mathcal{K}^{x'x}[A] =
    \int_{0}^{\infty} \!\ud T\, \e^{-im^2 T} \!
    \int_{x(0) = x}^{x(T) = x'}\hspace{-1.5em}\mathcal{D}x(\tau)\,
    \e^{iS[x(\tau),A]}\,\mathrm{Spin}[F]\,.
\ee
The spin factor, $\mathrm{Spin}[F]$ serves as an augmentation to the scalar propagator,~\eqref{eq:propagator_pos_pi}, whose action reappears in the same form here, to include the spin degrees of freedom along a trajectory:
\begin{equation}
    \mathrm{Spin}[F] = \mathcal{P} \e^{-\frac{ie}{2} \int_{0}^{T} d\tau \, F_{\mu\nu}(x(\tau)) \sigma^{\mu\nu}}\,,
\end{equation}
where $\sigma^{\mu\nu} = \frac{i}{2}[\gamma^{\mu}, \gamma^{\nu}]$ are the Lorentz generators for the spinor representation and $\mathcal{P}$ indicates path ordering of the matrix-valued potential along the trajectories. It has a path integral representation in terms of Grassmann fields
\pagebreak
\begin{align} 
    \label{eq:SpF}
     \mathrm{Spin}[F] &= 
    \mathrm{symb}^{-1} \mathfrak{W}[F]\,,\\
    \mathfrak{W}[F] &= 2^{-\frac{D}{2}}
    \oint_{\textrm{A/P}}\hspace{-0.75em}\mathcal{D}\psi(\tau) \,    
    \e^{i\int_{0}^{T}\ud \tau \bigl[\frac{i}{2}\psi\cdot\dot{\psi}
    + ie\psi_\eta(\tau) \cdot F(x(\tau)) \cdot\psi_\eta(\tau)\bigr]}\,.
    \label{eq:WF}
\end{align}
with anti-periodic boundary conditions (A/P) on the field, $\psi^{\mu}(\tau)$. The kernel is Dirac matrix-valued so as to produce the Dirac structure of the propagator; to account for that, another Grassmann variable has been introduced via $\psi_\eta(\tau)\coloneqq\psi(\tau)+\eta$, which is acted on by a symbolic map, converting Grassmann variables to anti-symmetrised products of Dirac matrices,
\begin{equation}
    \textrm{Symb}\Big\{ \gamma^{[\mu_{1}} \gamma^{\mu_{2} \cdots \gamma^{\mu_{n}]}} \Big\} = (-i\sqrt{2})^{n} \eta^{\mu_{1}}\eta^{\mu_{2}}\cdots \eta^{\mu_{n}} 
\end{equation}
(see~\cite{Fradkin:1991ci, Ahmadiniaz:2020wlm} and \cite{Copinger:2023ctz} for specifics as they pertain to our notations and (3+1)-dimensions). Similarly, we may also record the one-loop effective action as
\be
    \bar{\Gamma}[A] = \frac{i}{2}
    \int_{0}^{\infty}\frac{\ud T}{T} \,\e^{-im^{2} T}\,
    \int_{x(0) = x(T)} \hspace{-1.5em}\mathcal{D}x(\tau)\, 
    \e^{iS[x(\tau),A]}\,
    \mathrm{tr}\,\mathrm{Spin}[F]\;.
\ee
One advantage of this representation is that the scalar and spinor calculations are more closely related than in the standard formalism of their respective QFTs (e.g., by the inclusion of the spin factor and a Dirac trace/operator). Moreover, this treatment of the orbital and spin degrees of freedom naturally leads to a convenient ``spin-orbit decomposition'' outlined in \cite{Ahmadiniaz:2020wlm}.

As before, we may arrive at the $N$-photon dressed representation for both the propagator and effective action by decomposing the gauge field as $eA_\mu(x) = a_\mu(x) +eA_\mu^\gamma(x)$ and expanding to multi-linear order in $A_\mu^\gamma(x)=\sum_{i=1}^{N} \varepsilon_{i \mu} \e^{i k_{i} \cdot x}$. Employing the linearisation operator, we may similarly decompose the propagator into a ``leading'' and ``subleading'' part \cite{Ahmadiniaz:2020wlm}
\begin{align}
    \mathcal{S}^{x'x}_N[a] &= \big(-i\slashed{\partial}_{x'}+\slashed{a}(x') - m\big)\mathcal{K}^{x'x}_N[a]+
    \sum_{i = 1}^{N}e \slashed{\varepsilon}_{i}\e^{i k_{i}\cdot x'}\mathcal{K}^{x'x}_{N-1}[a]\;,\\
    \mathcal{K}^{x'x}_N[a] &=(-ie)^{N}
    \int_{0}^{\infty} \!\ud T \int\prod_{i=1}^{N}d\tau_i\,\e^{-im^2 T} \!
    \int_{x(0) = x}^{x(T) = x'}\hspace{-1.5em}\mathcal{D}x(\tau)\,
    \e^{iS[x(\tau),a]
    -i\int_{0}^{T}\!\ud\tau\,\mathcal{J}_0\cdot x(\tau)}
    \,\mathrm{Spin}_N[f]
    \Big|_{\textrm{lin}\,N}\;,
\end{align}
where, under the sum, in $\mathcal{K}_{N-1}$ we omit the $i^{\textrm{th}}$ photon. Here, the $N$-photon dressed spin factor is determined by 
\begin{align}
   \mathfrak{W}_N[f] &= 
    \e^{\sum_{i=1}^N\tfrac{\delta}{\delta \theta(\tau_i)}\cdot \tilde{f}_i \cdot
    \tfrac{\delta}{\delta \theta(\tau_i)}}
     \Psi_\theta[f]
    \Big|_{\theta=0}\,,\\
    \Psi_\theta[f] &=
    2^{-\frac{D}{2}}\oint_{\textrm{A/P}}\hspace{-0.75em}
    \mathcal{D}\psi(\tau) \,    
    \e^{i\int_{0}^{T}\ud \tau \bigl[\frac{i}{2}\psi\cdot\dot{\psi}
    + i\psi_\eta \cdot f \cdot\psi_\eta+i\theta\cdot \psi_\eta\bigr]}\,, 
\end{align}
with Grassmann source, $\theta(\tau)$ and we introduced the scattering photons' field strength tensors by $\tilde{f}_{\mu\nu}=k_\mu \varepsilon_\nu - k_\nu \varepsilon_\mu$. Likewise the N-photon dressed effective action can be found as
\be
    \bar{\Gamma}_N[a] = \frac{i}{2}(-ie)^{N}
    \int_{0}^{\infty}\frac{\ud T}{T} \int\prod_{i=1}^{N}d\tau_i\,\e^{-im^{2} T}\,
    \int_{x(0) = x(T)} \hspace{-1.5em}\mathcal{D}x(\tau)\, 
    \e^{iS[x(\tau),a]
    -i\int_{0}^{T}\!\ud\tau\,\mathcal{J}_0\cdot x(\tau)}\mathrm{tr}\,\mathrm{Spin}_N[f]
\Big|_{\textrm{lin}\,N}\;.
\ee

The above representations show that in the worldline formalism treating spinors at the level of the kernel and effective action amounts to a simple insertion of the spin factor. Moreover, for the non-null plane wave fields considered in this work, the appearance of the spin factor under the coordinate path integral does not inhibit the ``hidden Gaussianity'' ~\cite{Schubert:2023gsl} uncovered for the scalar case. However, unlike the scalar case whose $\rho^2$ dependence \emph{in a partial resummation} can be entirely characterised as a differential operator,~\eqref{eq:int_chi}, acting on the propagator or effective action, we now show that there is an additional $\rho^2$ dependence explicit in the background field strength. It should be noted that an exact solution for the spinor Green function on the worldline is available for this background~\cite{Copinger:2023ctz}; however it proves convenient to rather separate the $\rho^2$ dependence in the spin factor and evaluate the Grassmann path integral exactly about the plane wave portion (let us stress that an implicit non-null field dependence will remain the plane wave field strength). We write for the non-null field strength,
\begin{align}
    f_{\mu\nu}(x^{\tri})=\mathrm{f}_{\mu\nu}(x^{\tri}) + \frac{\rho^{2}}{4}\bar{\mathrm{f}}_{\mu\nu}(x^{\tri})
\end{align}
where now the plane wave part and the complementary part of the field strength are 
\begin{align}
    \mathrm{f}_{\mu\nu}(x^{\tri}) &\coloneqq n_\mu a'_\nu(x^{\tri})-n_\nu a'_\mu(x^{\tri})\,,
    \label{eq:f_pw_spin} \\
    \bar{\mathrm{f}}_{\mu\nu}(x^{\tri}) &= \bar{n}_\mu a'_\nu(x^{\tri})-\bar{n}_\nu a'_\mu(x^{\tri})\,.
    \label{eq:f_pwc_spin}
\end{align}
Such a decomposition leads to the following representation that expands about the plane :
\be
   \mathfrak{W}_N[f] = 
    \e^{\sum_{i=1}^N\tfrac{\delta}{\delta \theta(\tau_i)}\cdot \tilde{f}_i \cdot
    \tfrac{\delta}{\delta \theta(\tau_i)} - \frac{\rho^2}{4}\int^T_0 \ud \tau\,  \frac{\delta}{\delta\theta(\tau)} \cdot \bar{f}(\tau)\cdot \frac{\delta}{\delta\theta(\tau)}}
   \Psi_\theta[\mathrm{f}] \Big|_{\theta=0}\,, 
\ee
where $\rm{f}(\tau)$ denotes~\eqref{eq:f_pw_spin} but should be interpreted distinctly depending on whether the orbital path integral has / has not been completed. Before completion, $\textrm{f}(\tau)$ depends on the varying path integration variable ${\rm{f}}(x(\tau))$ under the integral; after the path integral over orbital degrees of freedom has been completed, this argument is simply replaced by the classical trajectory, $x_{\rm cl.}(\tau)$, given in e.g.~\eqref{implicit_a_line} on the line in coordinate space (or as is appropriate for loop or line in momentum space). From here, a reliable partial resummation of $\rho^2$ can be accomplished. 

The Grassmann path integral defining $\Psi_{\theta}[\textrm{f}]$ has a known exact solution~\cite{Copinger:2023ctz}, since $[\mathrm{f}(\tau),\mathrm{f}(\tau')]=0$ and $\mathrm{f}(\tau)\mathrm{f}(\tau')\mathrm{f}(\tau'')$ by virtue of $n^2=n\cdot a=0$ (note that even though we have $a^\LCperp(x^{\tri})$, this decomposition is still possible since $a$ is orthogonal to $x^{\tri}$). We find 
\be
    \Psi_\theta [\mathrm{f}] =
    \e^{-\int_{0}^{T}\!\ud\tau \, [\eta \cdot \mathrm{f}(\tau) \cdot \eta +{\j(\tau)} \cdot \eta]
    -\int_{0}^{T}\!\ud\tau
    \ud\tau'[\eta \cdot \mathrm{f}(\tau) \cdot \mathfrak{G}(\tau, \tau') \cdot \j(\tau')+\frac{1}{4}\j(\tau)\cdot \mathfrak{G}(\tau, \tau')\cdot \j(\tau')]}\,,
\ee
with the spinor Green function in the background reading
\be
\mathfrak{G}(\tau,\tau')=\mathrm{sgn}(\tau-\tau')
\Bigl[1-2\int_{\tau'}^{\tau}\!\ud\sigma \,\mathrm{f}(\sigma)
+2\Big(\int_{\tau'}^{\tau}\!\ud \sigma\, \mathrm{f}(\sigma)\Big)^{2}\Bigr]
+T\llangle \mathrm{f}\rrangle\Bigl[1-2\int_{\tau'}^{\tau}\!\ud \sigma\, \mathrm{f}(\sigma)\Bigr]\,,
\ee
where we have made use of a matrix representation for the spacetime indices.

Equipped with the above evaluation of the spin factor, we may immediately solve for the full perturbative in $\rho^2$ expressions for the non-null kernel of the $N-$photon dressed propagator and effective action. We focus our attention here to just the kernel, and this is because, as we will show below, the spinor amplitude only depends on the kernel in the worldline master formulae after truncation on both sides. Moreover, since the ``hidden'' Gaussianity still persists even with the spin factor, all the steps taken to arrive at the scalar propagator may be applied here as well, and we find that
\begin{align}
   \hspace{-1em} & \mathcal{K}_{N}^{x'x}[a] = i(-ie)^{N}\int_{0}^{\infty} dT \, (4\pi iT)^{-2}  \prod_{i = 1}^{N} \int_{0}^{T}d\tau_{i}\e^{i \rho^{2}\int \ud \tau' \ud \tau'' \Delta(\tau', \tau'') \frac{\delta}{\delta \xi(\tau')} \frac{\delta}{\delta \xi(\tau'')}} \Big\lbrace \e^{ -iM^{2}(a)T-i \frac{z^{2}}{4T} - i z \cdot \llangle a \rrangle + i\sum_{j=1}^{N} \big[ k_{j} \cdot (x + z \frac{\tau_{j}}{T}) - i \varepsilon_{j} \cdot \frac{z}{T} \big]} \notag \\
 \hspace{-1em} & \times \e^{ 2\sum_{j=1}^{N} \big[ \big(\llangle a \rrangle - a(\tau_{j})\big)\cdot \varepsilon_{j} - i I(\tau_{j})\cdot k_{j} \big]
-i\sum_{i,j=1}^{N}[\Delta_{ij}k_{i}\cdot k_{j} - 2i\ddel_{ij}\varepsilon_{i}\cdot k_{j} - \ddeld_{ij} \varepsilon_{i}\cdot\varepsilon_{j}]
} \,\mathrm{Spin}_N[f]\Big\rbrace \Big|_{\mathrm{lin}\,N}^{\xi(\tau) \rightarrow 2i\sum_{j=1}^{N} [\deld(\tau, \tau_{j}) \varepsilon_{j}^{\tri} + i\Delta(\tau, \tau_{j})k_{j}^{\tri}]}\,,
\label{Kxx_full}
\end{align}
in close analogy to the scalar case of~\eqref{Dxx_full}. The implicit dependence for the gauge field still remains the same as for the scalar case, as shown in~\eqref{implicit_a_line}, and here there is also an implicit dependence in $\mathrm{f}_{\mu\nu}$ of~\eqref{eq:f_pw_spin}. Likewise, the full $N-$photon dressed effective action may be immediately inferred from the scalar case, and we find that
\begin{align}
    &\bar{\Gamma}_{N}[a] = -\frac{1}{2}(-ie)^{N}\int_{0}^{\infty} \frac{dT}{T}(4\pi iT)^{-2}\prod_{i = 1}^{N} \int_{0}^{T}d\tau_{i} \int d^{4}\xnot\, \e^{\frac{i}{2} \rho^{2}\int \ud \tau' \ud \tau'' G(\tau', \tau'') \frac{\delta}{\delta \xi(\tau')} \frac{\delta}{\delta \xi(\tau'')}} \Big\lbrace \e^{-iM^{2}(a) T + i\sum_{j=1}^{N}  k_{j} \cdot \xnot} \notag \\
 &  \e^{2\sum_{j=1}^{N} \big[\big(\llangle a \rrangle - a(\tau_{j}) \big)\cdot\varepsilon_{j}+i\big( I(\tau_{j}) - \llangle I \rrangle \big)\cdot k_{j}\big] 
-\frac{i}{2} \sum_{i,j=1}^{N}[G_{ij}k_{i}\cdot k_{j} - 2i\Gd_{ij}\varepsilon_{i}\cdot k_{j} + \Gdd_{ij} \varepsilon_{i}\cdot\varepsilon_{j}]
} \,\mathrm{tr}\,\mathrm{Spin}_N[f] \Big\rbrace \Big|_{\textrm{lin}\,N}^{\xi(\tau) \rightarrow i\sum_{j=1}^{N} [\Gd(\tau, \tau_{j}) \varepsilon_{j}^{\tri} -iG(\tau, \tau_{j})k_{j}^{\tri}]}\,,
\label{effa_full_spin}
\end{align}
where likewise the implicit dependence here is given by~\eqref{implicit_a_loop}.  Taking the trace of the spin factor projects onto the sector with zero factors of the Grassmann variables, $\eta^{\mu}$ and produces a factor of 4 for that contribution (the trace of the antisymmetric products of $\gamma$-matrices spanned by the image of the inverse symbol map is zero). The above representations are exact in the non-null parameter, albeit formally so, and in the next section we carry out a perturbative expansion in the explicit dependence on $\rho^{2}$.

\subsection{Multi-photon scattering on the loop and line}

To arrive at an explicit representation for scattering amplitudes for spinor particles, we again employ a partial resummation with explicit expansion about small $\rho^2$ as was accomplished for scalars. For illustration we again truncate up to $\mathcal{O}(\rho^2)$, which for the orbital contributions is shown in~\eqref{eq:func_der_small}. There is, of course now an additional explicit truncation arising from the spin factor:
\begin{equation}
   \mathfrak{W}_N[f] = \Bigl\{ 1- \frac{\rho^2}{4}\int^T_0 \ud \tau\, \frac{\delta}{\delta\theta(\tau)} \cdot \bar{f} (\tau)\cdot \frac{\delta}{\delta\theta(\tau)} \Bigr\}
    \e^{\sum_{i=1}^N\tfrac{\delta}{\delta \theta(\tau_i)}\cdot \tilde{f}_i \cdot
    \tfrac{\delta}{\delta \theta(\tau_i)}}
   \Psi_\theta[\mathrm{f}] \Big|_{\theta=0} +\mathcal{O}(\rho^4)\,,
   \label{eq:WNf}
\end{equation}
where we recall that the spin factor here still must be acted upon by the functional derivatives with respect to $\xi(\tau)$ given in~\eqref{eq:func_der_small}. However, since we have truncated to $\mathcal{O}(\rho^4)$, it is immediate that the functional derivatives will not need to act on the $\rho^2$ term in the brackets to this order. 

We next turn our attention to the scattering of $N$ photons on the matter line for spinors. Let us start by noting that we will find that much like the plane-wave case~\cite{Copinger:2023ctz}, and shock-wave case~\cite{Copinger:2024twl}, the on-shell scattering amplitude will only depend on the spinor kernel, $\mathcal{K}_N[a]$. To see this, the scattering amplitude may be written down for the non-null background as
\begin{align}\label{M_N}
    \mathcal{M}_{Ns's}^{p'p} =  i 
    \lim_{p'^2,p^2\rightarrow m^2}
    \frac{1}{2m}\int d^{4}x'd^{4}x\, \e^{i\tilde{p}'\cdot x'-ip\cdot x}\,\bar{u}_{s'}(p')
    &(p^{\prime 2}-m^2)\Bigl\{ \Bigl[-1
    {+\frac{1}{2m} \delta \slashed{a}(x^{\prime +}))}
    \Bigr]\,\mathcal{K}_N^{x'x}\notag\\
    & \hspace{4em}+\frac{e}{2m}\sum_{i=1}^{N}\slashed{\varepsilon}_{i}\e^{ik_{i}\cdot x'}\,\mathcal{K}_{N-1}^{x'x}\Bigr\}(p^2-m^2)u_{s}(p)\,,
\end{align}
where we have made use of the free spinor on-shell solutions, e.g, $\bar{u}_{s'}(p')(\slashed{p}'+m)^{-1}=\bar{u}_{s'}(p')(2m)^{-1}$~\cite{Ahmadiniaz:2021gsd}, and likewise for $u_s(p)$, corresponding to spin 1/2 fermions with spin state $s$. The factors (1), with the $\delta\slashed{a}(x^{\prime \LCp})$ term and (2), with the summation over the $N-$photons are on the same level as ``subleading'' factors in the plane wave and non-null backgrounds. They vanish after LSZ-truncation on either side; we refer to the plane wave case~\cite{Copinger:2023ctz}~(see, for example, Eq. (81) and the surrounding discussion) for details as the asymptotic setup is essentially identical for the non-null case discussed here. We find that after factors (1) and (2) vanish, the LSZ-reduced amplitude becomes simply
\be
    \mathcal{M}_{Ns's}^{p'p} =  -i
    \lim_{p'^2,p^2\rightarrow m^2}\frac{1}{2m} \bar{u}_{s'}(p')(p^{\prime 2}-m^2)\mathcal{K}_N^{{\tilde{p}}'p} (p^2-m^2)u_{s}(p)\,,
\ee
where we can now see that one needs only to treat the kernel.

To arrive at the Fourier transformed kernel, one need only follow the steps as provided in Appendix ~\ref{sec:appendix_scalar} for the scalar propagator. The difference between the two amounts to the inclusion of the spin factor and the spin factor only contains implicit end-point, $x^{\tri}$ and $x^{\prime \tri}$, dependence. Therefore, all the steps taken for the Fourier transform of the scalar propagator apply here as well, and one may immediately write down the kernel in momentum space as
\begin{align}
    \mathcal{K}_{N}^{\tilde{p}'p}&=-i(-ie)^{N}\hat{\delta}_{-\LCperp}(p-K-\tilde{p}^{\prime})\int_{0}^{\infty}dT\,\e^{i(p'^{2}-m^{2}+i0^{+})T}\frac{1}{2}\int dx^{\tri}\e^{i(\tilde{p}^{\prime}+K-p)_{\tri}x^{\tri}}\prod_{i=1}^{N}\int_{0}^{T}d\tau_{i}\,[1+\rho^{2}\mathcal{I}_{N}]\\&\times \e^{-2iTp^{\prime}\cdot\llangle\delta a\rrangle+iT\llangle\delta a^{2}\rrangle-2\sum_{i=1}^{N}[a_{i}\cdot\varepsilon_{i}+i\int_{0}^{\tau_{i}}d\tau a(\tau)\cdot k_{i}]+i(2\tilde{p}^{\prime}+K)\cdot g+iS_N^{\textrm{BK}}}\mathrm{Spin}_{\rho^2N}[f]\Big|_{\mathrm{lin} N}\,.
    \notag
\end{align}
In this equation, we must determine the has explicit $\rho^2$ dependence of the spin factor, which is produced by the functional derivatives in \eqref{eq:WNf} \textit{and} by the functional derivatives with respect to $\xi(\tau)$ in \eqref{eq:func_der_small}. However, the latter act under the $\psi(\tau)$ path integral to simply pull down a derivative of the non-null field strength tensor sandwiched between two factors of $\psi_{\eta}(\tau)$, \textit{viz}.
\begin{align}
    \frac{\delta}{\delta \xi(\tau)} \e^{-\int_{0}^{T} d\tau' \, e \psi_{\eta}(\tau') \cdot f(\tau') \cdot \psi_{\eta}(\tau')} &= -e \psi_{\eta}(\tau) \cdot f'(\tau) \cdot \psi_{\eta}(\tau)\e^{-\int_{0}^{T} d\tau' \, e \psi_{\eta}(\tau') \cdot f(\tau') \cdot \psi_{\eta}(\tau')} \\
    &= -e \frac{\delta}{\delta \theta(\tau)} \cdot f'(\tau) \cdot \frac{\delta}{\delta \theta(\tau)}\e^{-\int_{0}^{T} d\tau' \, e \psi_{\eta}(\tau') \cdot f(\tau') \cdot \psi_{\eta}(\tau') - \int_{0}^{T}d\tau'\,  \theta(\tau') \psi_{\eta}(\tau')}\Big|_{\theta = 0}\,,
\end{align}
so these derivatives (acting on the spin factor) are expressible in terms of the functional derivatives, $\delta/\delta\theta(\tau)$. Putting these contributions together we find
\be
   \mathfrak{W}_{\rho^2N}[f] = \Big[ 1+ \rho^2\mathcal{I}_N^{\textrm{sp}} \Big]
    \e^{\sum_{i=1}^N\tfrac{\delta}{\delta \theta(\tau_i)}\cdot \tilde{f}_i \cdot
    \tfrac{\delta}{\delta \theta(\tau_i)}}
   \Psi_\theta[\mathrm{f}] \Big|_{\theta=0}\,,
   \label{eq:spin_final}
\ee
where, 
\begin{align}
    \mathcal{I}_N^{\textrm{sp}}&=\frac{i}{2}\int_{0}^{T}d\tau\int_{0}^{T}d\tau'|\tau-\tau'|\Bigl\{\frac{\delta}{\delta\theta(\tau)}\cdot f'(\tau)\cdot\frac{\delta}{\delta\theta(\tau)}\frac{\delta}{\delta\theta(\tau')}\cdot f'(\tau')\cdot\frac{\delta}{\delta\theta(\tau')}\label{eq:I_spin_final}\\
    &+4iP_N(\tau')\cdot a'(\tau')\frac{\delta}{\delta\theta(\tau)}\cdot f'(\tau)\cdot\frac{\delta}{\delta\theta(\tau)}\Bigr\}-\frac{1}{4}\int^T_{0} \ud \tau\, \frac{\delta}{\delta\theta(\tau)}\cdot \bar{f}(\tau)\cdot \frac{\delta}{\delta\theta(\tau)}\,.\notag
\end{align}
It is understood further that contributions like $\mathcal{I}_N\mathcal{I}_N^{\textrm{sp}}$ are omitted since they come with a factor of $\rho^4$. Last, after taking the functional derivatives, the implicit dependence in the arguments of the gauge field / field strength is the same as it was for the scalars,~\eqref{eq:implicit}.

All the steps involving truncation are identical to the scalar case, with the exception of the functional derivatives acting on the spin factor. Let us go ahead then and record the amplitude as
\begin{align}
    \mathcal{M}_{N}^{p'p}&=(-ie)^{N}\int d^{4}x\,\e^{i(K+\tilde{p}'-p)\cdot x}\int_{-\infty}^{\infty}\prod_{i=1}^{N}d\tau_{i}\,\delta\bigg(\sum_{j=1}^{N}\frac{\tau_{j}}{N}\bigg)\Big[1+\rho^{2}\mathcal{I}_{N}\Big]\e^{i({\tilde p}' +p)\cdot g+iS_N^{\textrm{BK}}}\\
    &\times \e^{-i\int_{-\infty}^{0}d\tau\, [2\tilde{p}'\cdot a(\tau)-a^{2}(\tau)]-i\int_{0}^{\infty}d\tau\, [2p'\cdot\delta a(\tau)-\delta a^{2}(\tau)]-2i\sum_{i=1}^{N}[\int_{-\infty}^{\tau_{i}}d\tau\, k_{i}\cdot a(\tau)-i\varepsilon_{i}\cdot a(\tau_{i})]}
     \frac{1}{2m} \bar{u}_{s'}(p')\mathrm{Spin}_{\rho^2N}[f]u_{s}(p) \Big|_{\mathrm{lin}N}\,.\notag
\end{align}
The only modification to the spin factor of~\eqref{eq:spin_final} and~\eqref{eq:I_spin_final} is the replacement $\int^T_0d\tau\to\int^\infty_{-\infty}d\tau$ for all propertime integrals. As in the scalar case, the implicit dependence also shifts as~\eqref{eq:implicit_LSZ}. So apart from the additional complications of evaluating the spin factor at the desired multiplicity $N$, the spinor amplitude is qualitatively similar to the scalar case,~\eqref{eq:scalar_amp_final}. As before with the scalar case, we take the opportunity to check the above truncation process on one side against known expressions for the spinor Volkov wavefunction; we reserve this discussion for the Appendix~\ref{sec:spin_wavefunction}.  

Let us finally turn our attention to the case of light-by-light scattering in QED in a non-null background for small $\rho^2$. Since the calcuation proceeds in close close similarity to the scalar case (and indeed for the spinor scattering on the line), we may immediately write down the expression for the one-loop photon scattering amplitude from~\eqref{effa_full_spin} as
\begin{align}
    \bar{\Gamma}_{N}[a] = -\frac{1}{4}(-ie)^{N}\hat{\delta}(K_{-})\hat{\delta}^{2}(\mathbf{K}_{\perp}) \int_{0}^{\infty} \frac{dT}{T} (4\pi i T)^{-2}\int_{-\infty}^{\infty} d\xnot^{\tri} 
    \,\e^{iK_{\tri}\xnot^{\tri}}\,\prod_{i=1}^{N}\int_{0}^{T}d\tau_{i} \, \e^{-i M^{2}(a) T} \,\mathrm{tr}\Big\lbrace \Big[ 1 + \rho^{2} \mathcal{J}_{N}\Big] \nonumber\\
    \times \e^{2\sum_{j=1}^{N} \big[\big(\llangle a \rrangle - a(\tau_{j}) \big)\cdot\varepsilon_{j}+i\big( I(\tau_{j}) - \llangle I \rrangle \big)\cdot k_{j}\big] 
-\frac{i}{2} \sum_{i,j=1}^{N}[G_{ij}k_{i}\cdot k_{j} - 2i\Gd_{ij}\varepsilon_{i}\cdot k_{j} + \Gdd_{ij} \varepsilon_{i}\cdot\varepsilon_{j}]
} \,\mathrm{Spin}_N[f] \Big|_{\textrm{lin}\,N} \Big\rbrace \,,
\label{eq:GammaNSpin}
\end{align}
where the spin factor takes the form of that on the line,~\eqref{eq:spin_final}, but with the  $\rho^2$ corrections now reading
\begin{align}\label{eq:I_loop_spin}
    \mathcal{I}_N^{\textrm{sp}}&=\frac{i}{2}\int_{0}^{T}d\tau\int_{0}^{T}d\tau'\,G(\tau,\tau')\Bigl\{\frac{\delta}{\delta\theta(\tau)}\cdot f'(\tau)\cdot\frac{\delta}{\delta\theta(\tau)}\frac{\delta}{\delta\theta(\tau')}\cdot f'(\tau')\cdot\frac{\delta}{\delta\theta(\tau')}\\
    &+4iQ_N(\tau)\cdot a'(\tau)\frac{\delta}{\delta\theta(\tau')}\cdot f'(\tau')\cdot\frac{\delta}{\delta\theta(\tau')}\Bigr\}-\frac{1}{4}\int^T_{0} \ud \tau\, \frac{\delta}{\delta\theta(\tau)} \cdot \bar{f}(\tau)\cdot \frac{\delta}{\delta\theta(\tau)}\,.\notag
\end{align}
And we finally remark that the same justifications for small $\rho^2$ as were employed for the scalar case apply here as well (see Sections \ref{sec:rho_exp_tree} and \ref{sec:rho_exp_loop}).

\subsection{Effective action -- non-null CCF}
For the case of a non-null CCF, the worldline path integral can be calculated exactly (again, to all orders in $\rho$). For the orbital degrees of freedom, we did this in Section \ref{sec:nnCCF}. The contribution from the spin factor, \eqref{eq:WF}, is calculated analogously to the worldline calculation presented in that section \cite{Ahmadiniaz:2017rrk, fppaper3}, by completing the square in $\psi$.

Evaluating the integral over $\psi$ in this way for the general case of open worldlines, we obtain a functional determinant and boundary term,
\begin{equation}
    \mathfrak{W}[F] = 2^{-\frac{D}{2}}\underset{\textrm{ABC}}{\textrm{Det}}{}^{\frac{1}{2}} \Big[ \partial_{\tau} + 2 e F\Big]\e^{-\eta \cdot F \cdot \eta + \e^{2} \eta \cdot F \cdot {}^{\circ}\!\mathcal{G}_{F}^{\circ} \cdot F \cdot \eta}\,,
\end{equation}
where ${}^{\circ}\!\mathcal{G}_{F}^{\circ} \equiv \int_{0}^{T}d\tau \int_{0}^{T} d\tau'\, \mathcal{G}_{F}(\tau, \tau')$ is the integrated fermionic worldline Green function in the background $F$. Since for us $F_{\mu\nu} = f_{\mu\nu} = a_{0}\big( \nn_{\mu}\delta_{\nu \perp} - \delta_{\mu \perp} \nn_{\nu}\big)$ is constant, we can calculate the Green function and its functional determinant as
\begin{align}
    \underset{\textrm{ABC}}{\textrm{Det}} \Big[ \partial_{\tau} + 2 e F\Big] &= 2^{\frac{D}{2}}\det\Big[\cosh(eFT)\Big]\\
    \mathcal{G}_{F}(\tau, \tau') &= \sigma(\tau - \tau') \frac{\e^{-ieFT \dot{G}_{B}(\tau, \tau')}}{\cos(eFT)} \\
    {}^{\circ}\!\mathcal{G}_{F}^{\circ} \ &= \frac{T^{2}}{(eFT)^{2}}\Big[eFT - \tanh(eFT)\Big]
\end{align}
so that
\begin{equation}
    \mathfrak{W}[F] = \det{}^{\frac{1}{2}} \big[ \cosh(eFT)\big]\e^{\eta \cdot \tanh(eFT) \cdot \eta}\,.
\end{equation}
We can then use the results \eqref{eq:feven} and \eqref{eq:fodd} to obtain
\begin{align}
    \cosh(eFT) &= \eta + \Big[ \frac{\cosh\big( e T \sqrt{ \rho^{2} a_{\perp}^{\prime 2}}\big) - 1}{ e T \sqrt{\rho^{2} a_{\perp}^{\prime 2}}} \Big]f^{2}\\
    \det\big[\cosh(eFT)\big] &= -\cosh^{2}\big[ e T \sqrt{\rho^{2} a_{\perp}^{\prime 2}}\big]\\
    \tanh(eFT) &= \frac{\tanh\big[ e T \sqrt{\rho^{2} a_{\perp}^{\prime 2}]}\big]}{\sqrt{\rho^{2} a_{\perp}^{\prime 2}}}f\,.
\end{align}
For the effective action, then, we take the trace, which, when combined with the orbital contribution, provides us with an effective Lagrangian 
\begin{align}
    \mathcal{L}(a) &= -\frac{1}{2}\int_{0}^{\infty} \frac{dT}{T} (4\pi i T)^{-\frac{D}{2}} \e^{-im^{2}T} \det{}^{-\frac{1}{2}} \Big[ \frac{\tanh(eFT)}{eFT} \Big]\\
    &= -\frac{1}{2}\int_{0}^{\infty} \frac{dT}{T} (4\pi i T)^{-\frac{D}{2}} \e^{-im^{2}T} \Big[\frac{\sqrt{2\rho^{2}} T a_{0}}{\tanh \big(\sqrt{2\rho^{2}}  T a_{0}\big)} - 1 -\frac{2}{3}  \rho^{2}T^{2}a_{0}^{2} \Big]\,,
    \label{eq:Laspin}
\end{align}
where we have again implemented renormalisation about $D = 4$, showing that the first order non-null correction corresponds to charge renormalisation. Note that in the electric-dominated case ($-4 \leqslant \rho^{2} < 0$) the poles remain in the same place on the imaginary time axis so, after Wick rotation, we obtain a corresponding Ritus-like formula,
\begin{equation}
    \Im \mathcal{L}^{(1)} = \frac{2\rho^{2}a_{0}^{2}}{8\pi^{3}} \sum_{n = 1}^{\infty} \frac{1}{n^{2}}\, \e^{- \frac{n \pi m^{2}}{\sqrt{2\rho^{2}} a_{0}}}\,.
\end{equation}

To obtain this from our expansion in the parameter $\rho^{2}$ we start from \eqref{eq:GammaNSpin} and restrict to $a_{\mu}(x^{\tri}) = a_{0}\delta_{\mu \perp} x^{\tri}$ evaluating on $N = 0$. The orbital and spin contributions do not mix at $\mathcal{O}(\rho^{2})$ (the first cross term would be $\rho^{4} \mathcal{J}_{N} \mathcal{I}_{N}^{\textrm{sp}}$). Since the field strength tensor is constant in this case, the pre-factor to the exponential in \eqref{eq:GammaNSpin} is
\begin{align}
  \textrm{tr}\Big\lbrace  \rho^{2}\Big[\mathcal{J}_{0} + \mathcal{I}_{0}^{\textrm{sp}}\Big] \Big\rbrace
\end{align}
where $\mathcal{J}_{0} = \frac{T^{2}}{6}a'(x^{\tri})^{2}$ as in \eqref{eq:J0}. From the spin factor, we obtain only
\begin{align}
    \mathcal{I}_{0}^{\textrm{sp}} &= \frac{1}{4} \textrm{tr} \int_{0}^{T}d\tau \, \frac{\delta}{\delta \theta(\tau)} \cdot \bar{f}_{\mu\nu}(x^{\tri}) \cdot \frac{\delta}{\delta \theta(\tau)} \Psi_{\theta}[f]\big|_{\theta=0}\\
    &= -\frac{1}{8} \int_{0}^{T}d\tau\, \bar{f}_{\mu\nu}(x^{\tri}) \mathfrak{G}^{\mu\nu}(\tau, \tau)\\
    &= -\frac{T^{2}}{8} \bar{f}_{\mu\nu} \llangle f^{\mu\nu}\rrangle
\end{align}
Again, with a constant field strength tensor this further reduces to
\begin{equation}
    \mathcal{I}_{0}^{\textrm{sp}} = -\frac{T^{2}}{2} a'(x^{\tri})^{2}\,. 
\end{equation}
Putting these together we obtain $\textrm{tr}\big\lbrace  \rho^{2}\big[\mathcal{J}_{0} + \mathcal{I}_{0}^{\textrm{sp}}\big] \big\rbrace = -2\big(\frac{1}{3} - 1\big)T^{2} \rho^{2} a'(x^{\tri})^{2}$ and evaluating the remaining exponential factors we arrive at
\begin{align}
    \bar{\Gamma}_{0}[a] \Big|_{\rho^{2}} / V^{4} = -\frac{1}{3}  \int_{0}^{\infty} \frac{dT}{T} (4\pi i T)^{-2} \e^{-im^{2}T} \rho^{2} T^{2} a_{0}^{2} \,,
\label{eq:Gamma0}
\end{align}
which agrees with the charge renormalisation contribution in \eqref{eq:Laspin}. Of course, higher order corrections could be determined in a straightforward algebraic way from the Master Formulae provided above -- and this is, as is known, well-adapted for implementation in computer algebra systems.

\section{Conclusions}\label{sect:conclusions}
In this work, we have developed a systematic framework for describing scattering processes in strong-field QED in the presence of non-null plane-fronted electromagnetic backgrounds, i.e., backgrounds whose wavefronts remain planar but whose phase direction satisfies $\nn^{2}\neq 0$. Such fields arise naturally as effective descriptions of laser propagation in dispersive media, for instance, plasmas, and therefore provide a controlled step beyond the idealised vacuum plane-wave backgrounds that underpin much of the existing strong-field QED literature.

Our central technical result is the construction of $N$-photon–dressed master formulae for both tree-level propagators and one-loop effective actions in scalar and spinor QED, valid to all orders in the background field and systematically expandable in the non-nullness parameter $\rho^{2}=\nn^{2}$. By exploiting the worldline path-integral formalism, we were able to retain the non-perturbative resummation characteristic of plane-wave (Volkov) physics while incorporating deviations from null propagation as controlled insertions acting on corresponding plane-wave-like expressions. In this way, the familiar plane-wave results emerge smoothly in the $\rho^{2}\to 0$ limit, while dispersive effects are captured order by order beyond it.

A key conceptual feature of our approach is that results in the non-null background can be expressed entirely in terms of classical solutions to the worldline equations of motion in \textit{vacuum}, a property shared with the case of exact plane-wave backgrounds. This structure naturally leads to both space-time and momentum-space representations of dressed propagators, as well as their LSZ-truncated amplitudes, thereby providing a unified description of scattering processes at both tree and loop level (see \eqref{eq:unified_tree_loop}). The fundamental reason behind this is the semi-classical exactness of the worldline path integral in our class of backgrounds.

As nontrivial checks on our framework, we have shown that known results are recovered in appropriate limits, including the exact plane-wave expressions and previously derived leading-order non-null corrected wavefunctions and scattering amplitudes. We also examined a special exactly solvable case of constant crossed fields with $\rho^{2}\neq 0$, illustrating how the non-null deformation leads to qualitative new physical effects -- most notably the appearance of a non-vanishing imaginary part of the effective action, reflecting the (known) fact that pair production may no longer be forbidden once the null condition is relaxed.

The formalism developed here is also well-suited to extending the by now long-running efforts to find closed-form expressions for the low energy limit of the $N$-photon scattering amplitudes (see \cite{Ahumada:2025dyc, Ahmadiniaz:2023jwd, MishaLow, Edwards:2018vjd, Martin:2003gb}). This limit corresponds to projecting onto the multi-linear contribution to the amplitudes in the external photon momenta, $k_{i}$, valid when their characteristic energy, $\omega_{i} / m \ll 1$, with $m$ the mass of the matter particles represented by the line or loop. This is significant because it generally allows for the remaining worldline parameter integrals to be computed in closed form.

The approach presented opens several natural directions for further investigation. On the phenomenological side, it provides a systematic tool for assessing the impact of medium-induced dispersion and non-vanishing field invariants on strong-field processes such as nonlinear Compton scattering and Breit-Wheeler pair production in laser–plasma environments. One might also use this formalism, as it stands, to determine (at least perturbatively in $\rho^{2}$) the one-loop vacuum polarisation in our class of non-null fields and test whether the 1 particle reducible contributions to processes (known to be present in constant fields but corresponding simply to an additional renormalisation in a pure plane wave \cite{Ahmadiniaz:2019nhk}) are relevant in a non-null background. On the theoretical side, it would be interesting to extend the present analysis to backgrounds with additional structure, such as slowly varying transverse profiles, finite-duration pulses, or focusing effects, using the plane-wave sector as an analytic core around which further deformations are organised. More broadly, our results contribute to an ongoing effort to bridge the gap between exactly solvable strong-field QED models and the increasingly realistic electromagnetic fields encountered in modern high-intensity experiments and those expected in the future.

Finally, there are no fundamental obstacles to going beyond one-loop order. The worldline formalism extends naturally to higher loop calculations \cite{UsRep, ChrisRev, Kors:1998ew,  Fliegner:1997ra} whilst retaining its favourable properties such as summing over sets of Feynman diagrams related by exchange of photon insertions around the loop. Here we would expect the worldline path integral to be computable in closed form, although the contributions arising from the expansion in $\rho^{2}$ will no doubt be more complicated than their tree-level or one-loop counterparts presented here.

\vfill

\begin{acknowledgments}	
We would like to thank Anton Ilderton for fruitful discussions that have led to the improvement of this work, and Ivan Ahumada for providing helpful comments on a draft manuscript. The authors are supported by the EPSRC Standard Grants  EP/X02413X/1 (PC, JPE) and EP/X024199/1 (KR), the STFC Consolidated Grant ST/X000494/1 ``Particle Theory at the Higgs Centre" (KR), and the WPI program ``Sustainability with Knotted Chiral Meta Matter (WPI-SKCM${}^2$)" at Hiroshima University (PC). PC would further like to acknowledge support from the Research Start-up Support Fund of WPI-SKCM$^2$.
\end{acknowledgments}

\pagebreak

\appendix
\section{Path integral over auxiliary variables}\label{app:chi_integral}

Here we outline a different approach to dealing with the path integral over the auxiliary variables $\chi(\tau)$ and $\xi(\tau)$ defined appearing in \eqref{eq:int_chi}, to conclude with \eqref{Dxx_full}. 

We start with the fact that ~\eqref{eq:int_chi} remains quadratic in $\chi$ and so the path integral over $\chi$ can still be computed in closed form to give
$\mathcal{N} \exp \big(-\frac{i}{4 \rho^{2}} \int_{0}^{T} d\tau \, \dot{\xi}(\tau)^{2}\big)$ where $\mathcal{N} = \int \mathcal{D} \chi(\tau) \, \exp\big(-i\rho^{2}\int\!\ud \tau \ud\tau' \,  \chi(\tau)\chi(\tau') \Delta(\tau, \tau')\big)$ is a normalisation constant that ensures that in the limit $\rho^{2} \rightarrow 0$ the right hand side of (\ref{eq:int_chi}) reduces to $\delta[\mathcal{I}(\tau)]$. The path integral measure is defined by the continuum limit by $\int \mathcal{D}\chi(\tau) = \lim_{N\to \infty} \prod_{k = 1}^{N} \int d\chi_{k} \frac{T}{N}$ with $\det\big[\Delta(\tau_{i}, \tau_{j})\big]$ scaling as $N^{-N}$. 

Now, to obtain \eqref{Dxx_full} from this exact path integral, one should linearise the exponent that is quadratic in the gauge field with respect to $\xi(\tau)$:
\begin{equation}
    \e^{-i \int_{0}^{T}d\tau \int_{0}^{T}d\tau'\, a(\tau) \cdot a(\tau') \ddeld(\tau, \tau') } = \e^{ \int_{0}^{T} \xi(\tau) \frac{\delta}{\delta \theta(\tau)} }  \e^{-i \int_{0}^{T}d\tau \int_{0}^{T}d\tau'\, a(\bullet +\theta) \cdot a(\bullet + \theta) \ddeld(\tau, \tau') }\Big|_{\theta = 0}
\end{equation}
and then rescale $\xi(\tau) \to \rho \xi(\tau)$. The integral over $\xi(\tau)$ can then be computed by completing the square using the worldline Green function and results in \eqref{Dxx_full}, with $\theta(\tau)$ playing the role of $\xi(\tau)$. However, an alternative approach to an expansion in $\rho$ could also be developed prior to integrating over $\xi(\tau)$ by implementing the rescaling only, leading to
\begin{equation}
    \int_{DBC} \hspace{-1em} \mathcal{D}q(\tau)\,
\e^{i\int_{0}^{T}\ud\tau\big[-\frac{\dot{q}^{2}}{4}-\delta\mathcal{J}\cdot q\big]} = (4\pi i T)^{-\frac{3}{2}}  \int \mathcal{D}\xi(\tau) \e^{i\int_{0}^{T}d\tau\,\big[ -\frac{\dot{\xi}^{2}}{4} \big]- i \int_{0}^{T}d\tau \int_{0}^{T}d\tau'\, a(\bullet +\rho \xi) \cdot a(\bullet + \rho \xi) \ddeld(\tau, \tau') }\,,
\end{equation}
and then expanding the gauge potential in the \textit{explicit} factors of $\rho$. Insertions of $\xi(\tau)$ under the path integral are then dealt with using Wick's theorem and the worldline Green function (with an odd number of insertions vanishing).

\section{Scalar propagator Fourier transform}
\label{sec:appendix_scalar}

In this Appendix, we present more details of the Fourier transform of the scalar propagator used in the main text, keeping explicit terms up to $\mathcal{O}(\rho^{2})$. This derivation complements the approach in Section~\ref{sect:momentum-propagator}, where the momentum-space form of the dressed propagator was derived in a complementary way. The steps closely parallel those appearing in the Fourier transform of the spinor kernel; we therefore emphasise the points at which $\rho^{2}$-dependent structures arise, since these are essential for understanding the resummation scheme employed throughout the paper.

We begin from the defining representation
\begin{equation}
    \mathcal{D}_{N}^{\tilde{p}'p}
    =
    \int d^{4}x'\, d^{4}x\,
    \e^{i\tilde{p}'\cdot x' - i p \cdot x}
    \mathcal{D}_{N}^{x'x},
\end{equation}
where $\mathcal{D}_{N}^{x'x}$ is the resummed propagator.
The integrals over the transverse and minus light-cone components,
$x^{\LCperp},x^{\prime\LCperp}$ and $x^{\LCm},x^{\prime\LCm}$, can be performed immediately. Schematically the relevant Gaussian and delta-function integrals evaluate to
\begin{align}
    \int dx^{\prime\LCperp}dx^{\LCperp}\,\cdots
    &=4\pi iT\,
    \hat{\delta}_{\LCperp}(p-K-\tilde{p}^{\prime})
    \,\e^{-iT[\tilde{p}^{\prime}+\frac{1}{T}g]^{\LCperp2}
    +iT\llangle a\rrangle^{2}
    -2iT(\tilde{p}^{\prime}+\frac{1}{T}g)\cdot\llangle a\rrangle},
    \\
    \int dx^{\prime-}dx^{-}\,\cdots
    &=
    \Bigl\{1-i\rho^{2}T\partial_{z^{\tri}}^{2}\Bigr\}
    \hat{\delta}\Bigl(\tilde{p}'_{-}-\frac{1}{4T}z^{\tri}+\frac{1}{T}g_{-}\Bigr)
    \hat{\delta}\Bigl(p_{-}-\frac{1}{4T}z^{\tri}-K_{-}+\frac{1}{T}g_{-}\Bigr).
\end{align}
where the ellipses represent the exponential factor relevant to each of the integrals above. Next, we can perform an integration by parts to formally move the $\partial^2_{z^{\triangle}}$ off the delta function to the exponential factor. In fact one finds that the $\partial_{z^{\tri}}$ then acts only on the implicit dependence inside the argument to the gauge field. This in turn allows us to write 
\begin{equation}
    \partial_{z^{\triangle}}\int d\tau\,a(\tau)=\frac{1}{T}\int d\tau\Bigl(-\frac{T}{2}+\tau\Bigr)\frac{\delta}{\delta\xi(\tau)}\int d\tau'\,a(\tau')\,,
    \label{eq:deltaa}
\end{equation}
where the implicit dependence in the argument to the gauge field is still given as~\eqref{implicit_a_line}. Now we may complete one of the delta functions, performing the $x^{\prime\LCm}$ integral, which turns the argument of the gauge field into
\begin{equation}\label{eq:a_impl_FT}
    a^{\mu}(\tau_i)=a^{\mu}\Bigl(x^{\tri}+g^{\triangle}+(\tilde{p}^{\prime\triangle}+p^{\triangle})\tau_{i}-\sum_{j=1}^{N}|\tau_{i}-\tau_{j}|k_{j}^{\triangle}+\xi(\tau_{i})-\rho^{2}[\tilde{p}_{\triangle}^{\prime}+p_{\triangle}-K_{\triangle}^{-}+\frac{1}{T}g^{-}]\tau_{i}\Bigr)\,,
\end{equation}
from which it is clear that an additional $\mathcal{O}(\rho^{2})$ contribution appears in the argument of the background field. Now we see in contrast to~\eqref{eq:deltaa}, we have

\begin{align}
    \partial_{x^{\tri}}\int d\tau\,a(\tau)=\int d\tau\frac{\delta}{\delta\xi(\tau)}\int d\tau'\,a(\tau')
\end{align}
which allows us to reorganise the expression entirely in terms of functional derivatives acting on $\xi(\tau)$. Expanding the $\rho^{2}$-dependent term in~\eqref{eq:a_impl_FT} and subsequently recasting all explicit $\mathcal{O}(\rho^{2})$ contributions in terms of $\delta/\delta\xi(\tau)$ then leads directly to the final form:
\begin{align}
    \mathcal{D}_{N}^{\tilde{p}'p}&=(-ie)^{N}\hat{\delta}_{-\LCperp}(p-K-\tilde{p}^{\prime})\int_{0}^{\infty}dT\,e^{i(p'^{2}-m^{2}+i0^{+})T}\frac{1}{2}\int dx^{\tri}\e^{i(\tilde{p}^{\prime}+K-p)_{\triangle}x^{\tri}}\prod_{i=1}^{N}\int_{0}^{T}d\tau_{i}\notag\\&\times\int\mathcal{D}\xi\,\delta[\xi]\Bigl\{1+i\rho^{2}\int_{0}^{T}d\tau\int_{0}^{T}d\tau'\frac{1}{2}|\tau-\tau'|\frac{\delta}{\delta\xi(\tau')}\frac{\delta}{\delta\xi(\tau)}\Bigr\}\notag\\&\times \e^{-2iTp^{\prime}\cdot\llangle\delta a\rrangle+i\int_{0}^{T}d\tau\,\delta a^{2}-2\sum_{i=1}^{N}[a_{i}\cdot\varepsilon_{i}+i\int_{0}^{\tau_{i}}d\tau a(\tau)\cdot k_{i}]+i(2\tilde{p}^{\prime}+K)\cdot g+iS_{N}^{\textrm{BK}}}\Big|_{\mathrm{lin}N}\,,
\end{align}
where now the remaining argument to the gauge field reads
\begin{equation}
    a^{\mu}(\tau_i)=a^{\mu}\Bigl(x^{\tri}+g^{\triangle}+(\tilde{p}^{\prime\triangle}+p^{\triangle})\tau_{i}-\sum_{j=1}^{N}|\tau_{i}-\tau_{j}|k_{j}^{\triangle}+\xi(\tau_{i})\Bigr)\,.
\end{equation}
Note that up to this point, the derivation is insensitive to the inclusion of additional background-dependent structures, such as the spin factor. Consequently, the same steps apply directly to the spinor kernel discussed in the main text.
\subsection{Rewriting in the main-text representation}
\label{app:scalar_altrep}

At this stage, the Fourier-transformed propagator is expressed in a form in which the $\rho^{2}$ dependence appears both explicitly and implicitly through the background-field argument. While this representation follows naturally from the Fourier transform, it differs in appearance from the form adopted in the main text, which was obtained via a different derivation. To make the equivalence manifest, we now show how the expression derived above can be rewritten into the representation used in the main text.

Using momentum conservation $(p-K-\tilde{p}^\prime)_\LCm=0$, the argument to the gauge field becomes
\begin{equation}
    a^{\mu}(\tau_{i})=a^{\mu}\Bigl(x^{\triangle}+2\tilde{p}^{\prime\triangle}\tau_{i}-\sum_{j=1}^{N}\tilde{\Delta}_{ij}k_{j}^{\triangle}+\rho^{2}[p-K-\tilde{p}^{\prime}]_{\triangle}\tau_{i}+\xi(\tau_{i})\Bigr)\,.
\end{equation}
Expanding the $\rho^2$ contribution in a power series as a functional derivative gives
\begin{align}
    \mathcal{D}_{N}^{\tilde{p}'p}&=(-ie)^{N}\hat{\delta}_{-\perp}(p-K-\tilde{p}^{\prime})\int_{0}^{\infty}dT\,\e^{i(p'^{2}-m^{2}+i0^{+})T}\frac{1}{2}\int dx^{\triangle}\e^{i(\tilde{p}^{\prime}+K-p)_{\triangle}x^{\triangle}}\prod_{i=1}^{N}\int_{0}^{T}d\tau_{i}\\&\times\int\mathcal{D}\xi\,\delta[\xi]\Bigl\{1+i\rho^{2}\int_{0}^{T}d\tau\int_{0}^{T}d\tau'\frac{1}{2}|\tau-\tau'|\frac{\delta}{\delta\xi(\tau')}\frac{\delta}{\delta\xi(\tau)}+\rho^{2}[p-K-\tilde{p}^{\prime}]_{\triangle}\int_{0}^{T}d\tau'\tau'\frac{\delta}{\delta\xi(\tau')}\Bigr\}\\&\times \e^{-2iTp^{\prime}\cdot\llangle\delta a\rrangle+i\int_{0}^{T}d\tau\,\delta a^{2}-2\sum_{i=1}^{N}[a_{i}\cdot\varepsilon_{i}+i\int_{0}^{\tau_{i}}d\tau a(\tau)\cdot k_{i}]+i(2\tilde{p}^{\prime}+K)\cdot g+iS_{N}^{\textrm{BK}}}\Big|_{\mathrm{lin}N}\,.
\end{align}
The factor $[p-K-\tilde{p}^{\prime}]_{\triangle}$ above can be generated from the exponential as $-i \partial_{x^{\tri}}$, then integrated by parts, and finally expressed as a functional derivative, $\partial_{x^{\tri}}\to\int^T_0d\tau\,\delta/\delta\xi(\tau)$. This yields the alternate representation
\begin{align}
    \mathcal{D}_{N}^{\tilde{p}'p}&=(-ie)^{N}\hat{\delta}_{-\perp}(p-K-\tilde{p}^{\prime})\int_{0}^{\infty}dT\,\e^{i(p'^{2}-m^{2}+i0^{+})T}\frac{1}{2}\int dx^{\triangle}\e^{i(\tilde{p}^{\prime}+K-p)_{\triangle}x^{\triangle}}\prod_{i=1}^{N}\int_{0}^{T}d\tau_{i}\notag\\&\times\int\mathcal{D}\xi\,\delta[\xi]\Bigl\{1+i\rho^{2}\int_{0}^{T}d\tau\int_{0}^{T}d\tau'\frac{1}{2}\tilde{\Delta}_{\tau\tau'}\frac{\delta}{\delta\xi(\tau')}\frac{\delta}{\delta\xi(\tau)}\Bigr\}\notag\\&\times \e^{-2iTp^{\prime}\cdot\llangle\delta a\rrangle+i\int_{0}^{T}d\tau\,\delta a^{2}-2\sum_{i=1}^{N}[a_{i}\cdot\varepsilon_{i}+i\int_{0}^{\tau_{i}}d\tau a(\tau)\cdot k_{i}]+i(2\tilde{p}^{\prime}+K)\cdot g+iS_{N}^{\textrm{BK}}}\Big|_{\mathrm{lin}N}\,,
\end{align}
which matches precisely with the representation employed in the main text.

\section{Cross-check}\label{app:crosscheck}
In this Appendix, we perform a series of non-trivial checks on the formalism developed in the main text. In particular, we verify that our worldline expressions reproduce known results obtained via the standard diagrammatic approach in the Furry picture. These checks are not required for the derivations presented earlier, but serve to validate both the resummation procedure and the treatment of the non-null deformation parameter $\rho^{2}$.

We focus on two complementary tests: the recovery of known wavefunctions in non-null plane-wave backgrounds, and the verification of nonlinear Compton scattering amplitudes at low photon multiplicity.

\subsection{Wavefunction}\label{app:wavefn_compton}

We begin by examining the matter wavefunction obtained from the LSZ truncation of the dressed propagator. As discussed in the main text, truncation on a single external leg is sufficient to extract the outgoing wavefunction. Explicitly, we show that the followng relation holds:
\begin{equation}
    \lim_{p'^2\rightarrow m^2-i0^\LCp}\int d^{4}x\, \e^{-ip\cdot x}\phi^{\text{out}}_{p^\prime}(x)=\lim_{p'^2\rightarrow m^2-i0^\LCp}-i(p^{\prime2}-m^{2}+i0^{+})\mathcal{D}_{0}^{\tilde{p}'p}=\lim_{p'^2\rightarrow m^2}K(\infty)\,
\end{equation}
where $\mathcal{D}_{0}^{\tilde{p}'p}$ denotes the zero-photon sector of the momentum-space propagator. Using~\eqref{eq:FT}, the right-hand side reduces to
\begin{equation}
    \int d^{4}x\,[1+\rho^{2}\mathcal{I}_{0}]\,\e^{i(\tilde{p}^{\prime}-p)\cdot x-i\frac{1}{p^{\prime\tri}+p^{\tri}}\int_{x^{\tri}}^{\infty}[2p'\cdot\delta a(\varphi)-\delta a^{2}(\varphi)]d\varphi}\,
\end{equation}
with 
\begin{equation}
    \mathcal{I}_0=-i\frac{1}{(p^{\prime\tri}+p^{\tri})^{3}}\Bigl\{[\delta a^{2}(x^{\tri})-2p^{\prime}\cdot\delta a(x^{\tri})]\int_{x^{\tri}}^{\infty}d\varphi[\delta a^{2}(\varphi)-2p^{\prime}\cdot\delta a(\varphi)]\\-\int_{x^{\tri}}^{\infty}d\varphi[\delta a^{2}(\varphi)-2p^{\prime}\cdot\delta a(\varphi)]^{2}\Bigr\}\,.
\end{equation}
Next, using the fact that $p^{\prime\tri}-p^{\tri} = \frac{\rho^{2}}{2} (p^{\prime-}-p^{-})$, and imposing momentum conservation $p_-^\prime = p_-$, we retain terms up to $\mathcal{O}(\rho^2)$ in accordance with the resummed scheme. After an integration by parts in the $p^{\prime-}-p^{-}$ term, we arrive at a Fourier representation for the outgoing wavefunction:
\begin{align}
    \int d^{4}x\, \e^{-ip\cdot x}\varphi_{p'}^{\text{out}}(x)&=\int d^{4}x\Bigl[1+\rho^{2}\frac{1}{(2p^{\prime\tri})^{2}}[\delta a^{2}(x^{\tri})-2p'\cdot\delta a(x^{\tri})]\notag\\&+i\rho^{2}\frac{1}{(2p^{\prime\tri})^{3}}\int_{x^{\tri}}^{\infty}d\varphi[\delta a^{2}(\varphi)-2p^{\prime}\cdot\delta a(\varphi)]^{2}\Bigr]\e^{i(\tilde{p}^{\prime}-p)\cdot x-i\frac{1}{2p^{\prime\tri}}\int_{x^{\tri}}^{\infty}[2p'\cdot\delta a(\varphi)-\delta a^{2}(\varphi)]\ud\varphi}\,.
    \label{eq:scalar_wavefunction_final}
\end{align}
The wavefunction under the integral agrees\footnote{While~\cite{Mackenroth:2018rtp,Mackenroth:2020pct} provides spinor wavefunctions, the corresponding scalar expressions follow by omitting factors proportional to the $\gamma^\mu$ matrices and the free spinors.} with the known resummed expression calculated using the standard formalism of QFT~\cite{Heinzl:2016kzb,Mackenroth:2018rtp,Mackenroth:2020pct}.

Analogously, the ingoing wavefunction can be obtained by shifting the integration variable $\tau \to \tau + T$, then $x^{\tri} \to x^{\tri} + (p' + p)^{\tri} T$, followed by truncation with respect to the ingoing momenta $p$.

\subsection{Non-linear Compton}

Let us next confirm the $N=1$ non-linear Compton scattering expression; here the amplitude reduces to
\begin{align}
    \mathcal{A}_{1}^{p'p}&=-ie\int\ud^{4}x\,\e^{i(k_{1}+\tilde{p}'-p)\cdot x}[1+\rho^{2}\mathcal{I}_{1}]\e^{i({\tilde{p}}'+p)\cdot g}\notag\\&\times\e^{-i\int_{-\infty}^{0} \, \ud\tau[2\tilde{p}'\cdot a(\tau)-a^{2}(\tau)]-i\int_{0}^{\infty}\ud\tau\, [2p'\cdot\delta a(\tau)-\delta a^{2}(\tau)]-2i[\int_{-\infty}^{0}\ud\tau\, k_{1}\cdot a(\tau)-i\varepsilon_{1}\cdot a(\tau_{1})]}\bigg|_{\mathrm{lin}\, 1}\,,
    \label{eq:A_1_def}
\end{align}
with implicit dependence on $rho^{2}$ through the argument to the gauge field, $a(\tau)=a(x^{\tri}+(p^{\prime\tri}+p^{\tri})\tau-k_{1}^{\tri}|\tau|)$. To proceed we split this argument into $\tau \lessgtr 0$. We may use a change of variables for either plus or minus propertime, as appropriate, to convert propertime dependent quantities to their spacetime equivalents. In this way one can determine the components present in $\mathcal{I}_1$ to be
\begin{align}
    &-2i\int_{-\infty}^{0}d\tau\int_{-\infty}^{0}d\tau'|\tau-\tau'|\{p-a(\tau)\}\cdot a'(\tau)\{p-a(\tau)\}\cdot a'(\tau')\\
    &\quad=-i\frac{1}{(2p^{\tri})^{3}}\Bigl\{\{a^{2}(x^{\tri})-2p\cdot a(x^{\tri})\}\int_{-\infty}^{x^{\tri}}d\varphi\{a^{2}(\varphi)-2p\cdot a(\varphi)\}-\int_{-\infty}^{x^{\tri}}d\varphi\{a^{2}(\varphi)-2p\cdot a(\varphi)\}^{2}\Bigr\}\,,\notag\\
    &-4i\int_{0}^{\infty}d\tau\int_{-\infty}^{0}d\tau'|\tau-\tau'|\{p'-\delta a(\tau)\}\cdot a'(\tau)\{p-a(\tau)\}\cdot a'(\tau')\\
    &\quad=-i\frac{1}{2p^{\prime\tri}}\frac{1}{2p^{\tri}}\Bigl\{-\{a^{2}(x^{\tri})-2p\cdot a(x^{\tri})\}\int_{0}^{\infty}d\tau\{\delta a^{2}(\tau)-2p^{\prime}\cdot\delta a(\tau)\}\notag\\
    &\qquad-\{\delta a^{2}(x^{\tri})-2p^{\prime}\cdot\delta a(x^{\tri})\}\int_{-\infty}^{0}d\tau'\{a^{2}(\tau')-2p\cdot a(\tau')\}\Bigr\}\,,\notag\\
    &-2i\int_{0}^{\infty}d\tau\int_{0}^{\infty}d\tau'|\tau-\tau'|\{p'-\delta a(\tau)\}\cdot a'(\tau)\{p'-\delta a(\tau)\}\cdot a'(\tau')\\
    &\quad=-i\frac{1}{(2p^{\prime\tri})^{2}}\Bigl\{\{\delta a^{2}(0)-2p^{\prime}\cdot\delta a(0)\}\int_{0}^{\infty}d\tau\{\delta a^{2}(\tau)-2p^{\prime}\cdot\delta a(\tau)\}-\int_{0}^{\infty}d\tau\{\delta a^{2}(\tau)-2p^{\prime}\cdot\delta a(\tau)\}^{2}\Bigr\}\,,\notag\\
    &4\varepsilon_{1}\cdot a'(x^{\tri})\Bigl\{\int_{-\infty}^{0}d\tau\tau\{p-a(\tau)\}\cdot a'(\tau)-\int_{0}^{\infty}d\tau\tau\{p'-\delta a(\tau)\}\cdot a'(\tau)\Bigr\}\\
    &\quad=2\varepsilon_{1}\cdot a'(x^{\tri})\Bigl\{\frac{1}{(2p^{\tri})^{2}}\int_{-\infty}^{x^{\tri}}d\varphi\{a^{2}(\varphi)-2p\cdot a(\varphi)\}-\frac{1}{(2p^{\prime\tri})^{2}}\int_{x^{\tri}}^{\infty}d\varphi\{\delta a^{2}(\varphi)-2p^{\prime}\cdot\delta a(\varphi)\}\Bigr\}\,.
    \notag
\end{align}
Next we expand out the $\mathcal{O}(\rho^2)$ parts after a redefinition to spacetime coordinates to find for both of the following factors:
\begin{align}
    &\e^{-i\int_{-\infty}^{0}[2\tilde{p}'\cdot a(\tau)-a^{2}(\tau)]\ud\tau-2i\int_{-\infty}^{0}k_{1}\cdot a(\tau)\ud\tau}\\
    &\quad=\Bigl[1+\rho^{2}\frac{1}{(2p^{\tri})^{2}}[a^{2}(x^{\tri})-2p\cdot a(x^{\tri})]+\rho^{2}\frac{1}{(2p^{\tri})^{2}}\int_{-\infty}^{x^{\tri}}[a^{2}(\varphi)-2p\cdot a(\varphi)]\ud\varphi\notag\\
    &\qquad\times\Bigl\{-i\frac{1}{2p^{\tri}}[2p\cdot a(x^{\tri})-a^{2}(x^{\tri})]+i\frac{1}{2p^{\prime\tri}}[2p'\cdot\delta a(x^{\tri})-\delta a^{2}(x^{\tri})]-2\varepsilon_{1}\cdot a'(x^{\tri})\Bigr\}\Bigr]\e^{-i\frac{1}{2p^{\tri}}\int_{-\infty}^{x^{\tri}}[2p\cdot a(\varphi)-a^{2}(\varphi)]\ud\varphi}\,,\notag\\
    &\e^{-i\int_{0}^{\infty}\![2p'\cdot\delta a(\tau)-\delta a^{2}(\tau)]\ud\tau}\\
    &\quad=\Bigl[1-\rho^{2}\frac{1}{(2p^{\prime\tri})^{2}}[2p'\cdot\delta a(x^{\tri})-\delta a^{2}(x^{\tri})]+\rho^{2}\frac{1}{(2p^{\prime\tri})^{2}}\int_{x^{\tri}}^{\infty}\![2p'\cdot\delta a(\varphi)-\delta a^{2}(\varphi)]\ud\varphi\notag\\
    &\qquad\times\Bigl\{-i\frac{1}{2p^{\tri}}[2p\cdot a(x^{\tri})-a^{2}(x^{\tri})]+i\frac{1}{2p^{\prime\tri}}[2p'\cdot\delta a(x^{\tri})-\delta a^{2}(x^{\tri})]-2\varepsilon_{1}\cdot a'(x^{\tri})\Bigr\}\Bigl]\e^{-i\frac{1}{2p^{\prime\tri}}\int_{x^{\tri}}^{\infty}\![2p'\cdot\delta a(\varphi)-\delta a^{2}(\varphi)]\ud\varphi}\,.\notag
\end{align}
We have performed an integration by parts to arrive at the expressions on the RHS, where it is assumed that the above factors are present under the integral in~\eqref{eq:A_1_def}. Then gathering everything together, our representation of the $N-$photon dressed non-null amplitude agrees with the known diagrammatic expression (see~\cite{Mackenroth:2018rtp,Mackenroth:2020pct} for the spinor case in non-null fields) using standard techniques:
\begin{align}
    \mathcal{A}_{1}^{p'p}=&-ie\int\ud^{4}x\,e^{i(k_{1}+\tilde{p}'-p)\cdot x}\varepsilon_{1}\cdot(\tilde{p}^{\prime}+p-2a(x^{\tri}))\\
    &\quad\times\Bigl[1+\rho^{2}\Bigl\{\frac{1}{(2p^{\tri})^{2}}[a^{2}(x^{\tri})-2p\cdot a(x^{\tri})]-\frac{1}{(2p^{\prime\tri})^{2}}[2p'\cdot\delta a(x^{\tri})-\delta a^{2}(x^{\tri})]\notag\\
    &\qquad+i\frac{1}{(2p^{\tri})^{3}}\int_{-\infty}^{x^{\tri}}d\varphi\{a^{2}(\varphi)-2p\cdot a(\varphi)\}^{2}+i\frac{1}{(2p^{\prime\tri})^{3}}\int_{x^{\tri}}^{\infty}d\varphi\{\delta a^{2}(\varphi)-2p^{\prime}\cdot\delta a(\varphi)\}^{2}\Bigr\}\Bigr]\notag\\
    &\quad\times\e^{-i\frac{1}{2p^{\tri}}\int_{-\infty}^{x^{\tri}}[2p\cdot a(\varphi)-a^{2}(\varphi)]\ud\varphi-i\frac{1}{2p^{\prime\tri}}\int_{x^{\tri}}^{\infty}\![2p'\cdot\delta a(\varphi)-\delta a^{2}(\varphi)]\ud\varphi}\notag\\
    =&-ie\int\ud^{4}x\,[\tilde{p}'+p-2a(x^{\tri})]\cdot\varepsilon_{1}\e^{ik_{1}\cdot x}\varphi_{p'}^{\text{out}}(x)\varphi_{p}^{\text{{\rm in}}}(x)\,.
\end{align}

The checks presented here serve a dual purpose. First, they confirm that the master formulae derived in the main text correctly interpolate between the plane-wave limit and its non-null deformation. Second, it highlights the efficiency of the worldline approach: while the diagrammatic derivation requires careful bookkeeping of multiple contributions, the worldline representation yields a compact, ready-to-use expression at an early stage.

\subsection{Spinor Wavefunction}
\label{sec:spin_wavefunction}

Having shown the form of the wavefunction and non-linear Compton scattering for the scalar theory, let us consider the wavefunction for the spinor theory. Many of the steps here are similar to those above, therefore we focus on spin specific quantities in what follows. As a check of our formalism, let us confirm that it reproduces the known expression for the Volkov spinor wavefunction in a non-null background~\cite{Heinzl:2016kzb,Mackenroth:2018rtp,Mackenroth:2020pct}. For the wavefunction we need not truncate from both side but just one -- we compute the out wavefunction, truncating at positive asymptotic infinity. Let us first write down the wavefunction as
\begin{align}
    \phi_{p',s}^{\text{out}}(x)&=-\int d^{4}x\, \e^{i\tilde{p}'\cdot x}\bar{u}_{s}(p')(\slashed{p}'-m)\mathcal{S}_{0}^{x'x}\\
    &=-\int d^{4}x\, \e^{i\tilde{p}'\cdot x}\bar{u}_{s}(p')(\slashed{p}'-m)(\slashed{p}'-\delta\slashed{a}(x^{+})+m)\int_{0^{+}}^{\infty}dT\, \mathcal{K}_{0}^{x'x}\,.
\end{align}
Then following the exact same steps as used to reduce the amplitude, namely that the prefactor term proportional to $\delta\slashed{a}(x^\LCp)$ will vanish after truncation, we can determine the Fourier transformed wavefunction becomes
\begin{equation}
    \lim_{p'^2\rightarrow m^2-i0^\LCp}\int d^{4}x\, \e^{-ip\cdot x}\phi^{\text{out}}_{p^\prime,s}(x)=\lim_{p'^2\rightarrow m^2-i0^\LCp}-\bar{u}_{s}(p')(p^{\prime2}-m^{2}+i0^{+})\mathcal{K}_{0}^{\tilde{p}'p}\,,
    \label{eq:spin_wavefunc}
\end{equation}
and like the amplitude one need only consider the Fourier transformed kernel.

Following the same truncation process for the scalar wavefunction, after truncation, we find that the limits of propertime integration become $\int^\infty_0d\tau$ and the argument to the gauge field becomes $a(\tau)=a(x^{\tri}+(p'+p)^{\tri}\tau)$. The major difference computationally-wise compared to the scalar wavefunction is the evaluation of the spin factor, which we treat now. The $\mathcal{O}(\rho^{2})$ contribution from \eqref{eq:I_loop_spin} becomes for $N=0$
\begin{align}
    \mathcal{I}_0^{\textrm{sp}}
   & =\frac{2}{(p'+p)^{\tri 3}}(\delta a_{x}^{2}-2p'\cdot\delta a_{x})\eta^{+}\delta a_{x}\cdot\eta-\frac{4}{(p'+p)^{\tri 3}}\int_{0}^{T}d\tau(\delta a(\tau)^{2}-2p'\cdot\delta a(\tau))\eta^{+}\dot{a}(\tau)\cdot\eta\notag \\
    &-\frac{1}{(p'+p)^{\tri 2}}\frac{1}{2}\delta a_{x}^{2}\Bigl(1+\frac{2}{(p'+p)^{\tri}}\eta^{+}\delta a_{x}\cdot\eta\Bigr)-\frac{1}{4}\int_{0}^{T}d\tau\,\eta\, \mathfrak{G}^{\mathbb{T}}(\tau)\tilde{f}(\tau)\mathfrak{G}(\tau)\eta \, \e^{2\eta^{+}\delta a(x^{\tri})\cdot\eta}\,,
\end{align}
where 
\begin{align}
    \mathfrak{G}(\tau)_{\mu\nu}&=\Bigl\{\boldsymbol{1}-\int_{0}^{\infty}d\tau'f(\tau')\Big(\mathrm{sgn}(\tau-\tau')\Bigl[1-2\int_{\tau'}^{\tau}\!d\sigma\,f(\sigma)\Bigr]+\int_{0}^{\infty}d\sigma f(\sigma)\Big)\Bigr\}_{\mu\nu} \\
    &=\eta_{\mu\nu}+\frac{1}{(p'+p)^{\tri}}n_{\mu}(\delta a_{x}-2\delta a(\tau))_{\nu}-\frac{1}{(p'+p)^{\tri}}(\delta a_{x}-2\delta a(\tau))_{\mu}n_{\nu}+2\frac{[\delta a_{x}-\delta a(\tau)]\cdot\delta a(\tau)}{(p'+p)^{\tri 2}}n_{\mu}n_{\nu}\,.
\end{align}
Then after a few manipulations we can recast this in the form
\begin{equation}
    \int_{0}^{T}d\tau\, \eta\mathfrak{G}^{\mathbb{T}}(\tau)\tilde{f}(\tau)\mathfrak{G}(\tau)\eta=-2\eta^{-}\delta a_{x}\cdot\eta+16\int_{0}^{\infty}d\tau[\delta a^{2}(\tau)-\delta a_{x}\cdot\delta a(\tau)]a'\cdot\eta\, \eta^{+}\,.
\end{equation}
At this point in order to make the connection with the expression for the wavefunction in~\cite{Cronstrom,Becker-non-null} let us adopt a similar gauge choice of $a_\mu(x^\LCp)=\epsilon_{L\,\mu}a(x^\LCp)$ for spacetime independent laser polarisation $\epsilon_{L\,\mu}$. With this choice we obtain
\begin{equation}
\mathcal{I}_0^{\textrm{sp}}
    =\frac{4}{(p'+p)^{\triangle3}}(\delta a_{x}^{2}-2p'\cdot\delta a_{x})\eta^{+}\delta a_{x}\cdot\eta-\frac{1}{(p'+p)^{\triangle2}}\frac{1}{2}\delta a_{x}^{2}\Bigl(1+\frac{2}{(p'+p)^{\triangle}}\eta^{+}\delta a_{x}\cdot\eta\Bigr)+\frac{1}{2(p'+p)^{\triangle}}\eta^{-}\delta a_{x}\cdot\eta\,.
\end{equation}
Carrying out the remaining manipulations that are the same as those done for the orbital degrees of freedom, we can find the right side of~\eqref{eq:spin_wavefunc} becomes
\begin{equation}
\bar{u}_{s}(p')\mathrm{symb}^{-1}\int d^{4}x\Bigl[(1+\rho^{2}\mathcal{I}_{0})\Bigl(1+\frac{2}{(p'+p)^{\triangle}}\eta^{+}\delta a_{x}\cdot\eta\Bigr)+\rho^{2}\mathcal{I}_{0}^{\textrm{sp}}\Bigr]e^{i(\tilde{p}^{\prime}-p)\cdot x-i\frac{1}{p^{\prime\tri}+p^{\tri}}\int_{x^{\tri}}^{\infty}[2p'\cdot\delta a(\varphi)-\delta a^{2}(\varphi)]d\varphi}\,.
\end{equation}
Then using $p^{\prime\tri}-p^{\tri}=\frac{\rho^{2}}{2}(p^{\prime-}-p^{-})$ and expanding the $\rho^2$ contributions, and then rewriting them in terms of partial derivative w.r.t. $x^{\tri}$, we can finally confirm the known spacetime representation for the Volkov spinor wavefunction,
\begin{align}
    \int d^{4}x\, \e^{-ip\cdot x}\phi^{\text{out}}_{p^\prime,s}(x)&=\bar{u}_{s}(p')\int d^{4}x\Biggl\{\Bigl(1+\rho^{2}\frac{1}{(2p^{\prime\tri})^{2}}[\delta a^{2}(x^{\tri})-2p'\cdot\delta a(x^{\tri})]\\&+i\rho^{2}\frac{1}{(2p^{\prime\tri})^{3}}\int_{x^{\tri}}^{\infty}d\varphi[\delta a^{2}(\varphi)-2p^{\prime}\cdot\delta a(\varphi)]^{2}-\frac{\rho^2}{[2p^{\prime\triangle}]^{2}}\frac{1}{2}\delta a^{2}(x^{+})\Bigr)\Bigl\{1+\frac{1}{2p^{\prime\triangle}}\delta\slashed{a}_{x}\gamma^{\triangle}\Bigr\}\notag\\
    & +\frac{\rho^2}{(2p^{\prime\triangle})^{3}}\Bigl\{\delta a_{x}^{2}-2p^{\prime}\cdot\delta a_{x}\Bigr\}\delta\slashed{a}_{x}\gamma^{\triangle}-i\frac{\rho^2}{(2p^{\prime+})^{2}}\slashed{a}_{x}^{\prime}\gamma^{\triangle}
    \Biggr\}\, \e^{i(\tilde{p}^{\prime}-p)\cdot x-i\frac{1}{2p^{\prime\tri}}\int_{x^{\tri}}^{\infty}[2p'\cdot\delta a(\varphi)-\delta a^{2}(\varphi)]\ud\varphi}\,,\notag
\end{align}
in agreement with the known expression~\cite{Heinzl:2016kzb,Mackenroth:2018rtp,Mackenroth:2020pct}. Note that in addition to the explicit and implicit $\rho^2$ dependent terms there is also one term containing $\gamma^{\tri}$ that is unpaired with a $\rho^2$ factor that is indeed too proportional to $\rho^2$.

\section{Action of Quadratic Differential Operators on Gaussian Functions}
\label{app:quadratic_operator_gaussian}

In this Appendix, we derive a useful identity for the action of the exponential of a quadratic differential operator on a Gaussian function. The functional generalisation of this result is used to arrive at \eqref{eq:Gamma_as_det}.

\subsection{The result}

Let \(A\) and \(B\) be symmetric, positive-definite matrices. Then
\begin{equation}
\e^{\frac{1}{2}A_{ij}\partial_i \partial_j}
\Big(\e^{-x^{T}\cdot  B\cdot  x}\Big)\Big|_{x=0}
=
\Big[\det\!\left(
\frac{A^{-1}+2B}{A^{-1}}
\right)
\Big]^{-1/2}
\label{eq:quadratic_operator_identity}
\end{equation}

\subsection{Proof}

We start with the `Fourier representation' of the Gaussian operator:
\begin{align}
   \exp\Big(\frac{1}{2}A_{ij}\partial_i \partial_j\Big)\,= \frac{1}{\sqrt{\det(2\pi A) }} \int \mathcal{D}q \,
\exp\!\Big(
-\frac{1}{2} q^{T} A^{-1} q \Big) \exp\left(q_i\partial_i\right)\,.
\end{align}
Now applying the above operator on to $f(x)=\e^{-x^{T} B x}$ and using $\e^{q_i\partial_i}f(x)=f(x+q)$ leads to
\begin{align}
e^{\frac{1}{2}A_{ij}\partial_i \partial_j}
\Big(\e^{-x^{T} B x}\Big)
&=
\frac{1}{\sqrt{\det(2\pi A)}}
\int \mathcal{D}q \,
\exp\!\Big(
-\frac{1}{2} q^{T} A^{-1} q
-(x+q)^{T} B (x+q)
\Big).
\end{align}

Evaluating this expression at \(x=0\) yields
\begin{align}
\e^{\frac{1}{2}A_{ij}\partial_i \partial_j}
\Big(\e^{-x^{T} B x}\Big)\Big|_{x=0}
&=
\frac{1}{\sqrt{\det(2\pi A)}}
\int \mathcal{D}q \,
\exp\!\Big(
-\frac{1}{2} q^{T} (A^{-1}+2B) q
\Big).
\end{align}

The remaining integral is a standard Gaussian integral, giving
\begin{equation}
\int \mathcal{D}q \,
\exp\!\Big(
-\frac{1}{2} q^{T} (A^{-1}+2B) q
\Big)
=
\Big[\det\Big(\frac{2\pi}{A^{-1}+2B}\Big)\Big]^{1/2}.
\end{equation}

Combining terms, we arrive at
\begin{equation}
\e^{\frac{1}{2}A_{ij}\partial_i \partial_j}
\Big(\e^{-x^{T} B x}\Big)\Big|_{x=0}
=
\Big[\det\!\Big(
\frac{A^{-1}+2B}{A^{-1}}
\Big)
\Big]^{-1/2},
\end{equation}
which completes the proof.

\smallskip
\smallskip

\bibliography{BeyondBib}
\end{document}